\documentclass[11pt]{article}

\usepackage{amssymb}
\usepackage{graphicx}
\usepackage{layout}
\usepackage{amssymb}

\newcounter{CiUno}
\newcounter{CiDue}
\newcounter{CiTre}
\newcounter{CiQuattro}
\newcounter{CiCinque}
\newcounter{CiSei}
\newcounter{CiSette}
\newcounter{CiOtto}
\newcounter{CiNove}
\newcounter{CiDieci}
\newcounter{CiUndici}
\newcounter{CiDodici}
\newcounter{CiTredici}
\newcounter{CiQuattordici}

\newcommand{\fnref}[1]{~\ref{#1}}

\begin{document}

\title{The relativistic Sagnac effect:\\ two derivations}
\author{Guido Rizzi$^{\S,\P}$  and Matteo Luca Ruggiero$^{\S,\P}$\
\\
\small
%%EndAName
$^\S$ Dipartimento di Fisica, Politecnico di Torino,\\
\small $^\P$ INFN, Sezione di Torino\\
\small E-mail guido.rizzi@polito.it, matteo.ruggiero@polito.it}
\maketitle

\begin{abstract}
The phase shift due to the Sagnac Effect, for relativistic matter
and electromagnetic beams, counter-propagating in a rotating
interferometer, is deduced using two different approaches. From
one hand, we show that the relativistic law of velocity addition
leads to the well known  Sagnac time difference, which is the same
independently of the physical nature of the interfering beams,
evidencing in this way the universality of the effect. Another
derivation is based on a formal  analogy with the phase shift
induced by the magnetic potential for charged particles travelling
in a region where a constant vector potential is present: this is
the so called Aharonov-Bohm effect. Both derivations are carried
out in a fully relativistic context, using a suitable 1+3
splitting that allows us to recognize and define the space where
electromagnetic and matter waves propagate: this is an extended
3-space, which we call  \textit{relative space}. It is recognized
as the only space having an actual physical meaning from an
operational point of view, and it is identified as the 'physical
space of the rotating platform': the geometry of this space turns
out to be non Euclidean, according to  Einstein's early intuition.
\end{abstract}

\section{Introduction}\label{sec:intro}

The effects of rotation on space-time have always been sources of
stimulating and fascinating physical issues for the last
centuries. Indeed, even before the introduction of the concept of
space-time continuum, the peculiarity of the rotation of the
reference frame was recognized and understood. A beautiful example
is the Foucault's pendulum, which shows, in the context of
Newtonian physics, the absolute character of rotation.

At the dawn of the modern  scientific era, the notions of absolute
space and time, which are fundamental in the formulation of
classical laws of physics, were criticized by
Leibniz\cite{leibniz} and Berkeley\cite{berkeley}; consequently,
the concepts of absolute motion, and hence, of absolute rotation,
were questioned too. Mach's\cite{mach} analysis of the relativity
of motions revived the debate at the dawn of Theory of Relativity.
As it is well known, Mach's ideas played an important role and
influenced Einstein's approach. However, the peculiarity of
rotation, which is inherited by Newtonian physics, leads to
bewildering  problems even in the relativistic context. Actually,
after the publication of Einstein's theory, those who were
prejudicely contrary to Relativity found, in the relativistic
approach to rotation, important arguments against the
self-consistency of the theory. In 1909 an apparent logical
contradiction in the Special Theory of Relativity (SRT), applied
to the case of a rotating disk,
was pointed out by Ehrenfest\cite{ehrenfest}. Subsequently, in 1913, Sagnac%
\cite{sagnac13} evidenced an apparent contradiction in the SRT
with respect to experiments performed with rotating
interferometers: according to him

\begin{quote}
L'effet interf\'{e}rentiel observ\'{e} [...] manifeste directment
l'existence de l'\'{e}ther, support n\'{e}cessaire des ondes
lumineuses de Huygens et de Fresnel.
\end{quote}

Since those years, both the so-called 'Ehrenfest's paradox' and
the theoretical interpretation of the Sagnac effect have become
topical arguments of a discussion on the foundations of the SRT,
which is not closed yet, as the number of recent contributions
confirms.

We studied elsewhere\cite{rizzi02} the Ehrenfest's paradox, and we
showed that it can be solved on the basis of purely kinematical
arguments in the SRT. In this paper we are concerned with the
Sagnac effect: we are going to show that it can be  completely
explained in the SRT. To this end, we are going to give two
derivations of the effect.

On the one hand, using relativistic kinematics and, namely, the
law of velocity  addition, we are going to provide a ''direct''
derivation of the effect. In particular, the universality of the
effect - i.e. its independence from the physical nature and the
velocities (relative to the turntable) of the interfering beams -
will be explained.

On the other hand, we are going to give a "derivation by analogy"
which generalizes a previous work written by
Sakurai\cite{sakurai80}. Indeed, Sakurai outlined a beautiful and
far-reaching analogy between the Sagnac effect and the
Aharonov-Bohm effect\cite{aharonovbohm59}, and obtained a first
order approximation of the Sagnac effect. By generalizing
Sakurai's result, we shall obtain the Sagnac effect in full
theory, without any approximation, evidencing that the analogy
holds also in a fully relativistic context. To this end, we shall
use  Cattaneo's 1+3 splitting \cite{cattaneo}, \cite{catt1},
\cite{catt2}, \cite{catt3}, \cite{catt4}, that will enable us to
describe the geometrodynamics of the rotating frame in a simple
and powerful way: in particular, some Newtonian elements used by
Sakurai will be generalized to a relativistic context.\newline

The present paper is organized as follows: in Section
\ref{sec:history} a historical review of the Sagnac effect is
made; in Section \ref{sec:direct} the direct derivation is given,
while the derivation by analogy is outlined in  Section
\ref{sec:sagnacab}. Finally,  a thorough exposition of the
foundations of  Cattaneo's splitting is given in Appendix.

\section{A little historical review of the Sagnac effect}\label{sec:history}

\subsection{The early years}\label{ssec:early}

The history of the interferometric detection of the effects of
rotation dates back to the end of the XIX century when, still in
the context of the ether theory, Sir Oliver Lodge\cite{lodge93}
proposed to use a large interferometer to detect the rotation of
the Earth. Subsequently\cite {lodge97} he proposed to use an
interferometer rotating on a turntable in order to reveal the
rotation effects with respect to the laboratory frame. A detailed
description of these early works can be found in the paper by
Anderson\textit{\ et al.}\cite{anderson94}, where the study of
rotating interferometers is analyzed in a historical perspective.
In 1913 Sagnac\cite {sagnac13} verified his early
predictions\cite{sagnac05}, using a rapidly rotating light-optical
interferometer. In fact, on the ground of classical physics, he
predicted the following fringe shift (with respect to the
interference pattern when the device is at rest), for
monochromatic light waves in vacuum, counter-propagating along a
closed path in a rotating interferometer:
\begin{equation}
\Delta z=\frac{4\mathbf{\Omega \cdot S}}{\lambda c}
\label{eq:sagnac1}
\end{equation}
where $\mathbf{\Omega }$ is the (constant) angular velocity vector
of the turntable, $\mathbf{S}$ is the vector associated to the
area enclosed by the light path, and $\lambda $ is the wavelength
of light in vacuum (as seen in the local co-moving inertial frame
of the light source). The time difference associated to the fringe
shift (\ref{eq:sagnac1}) turns out to be
\begin{equation}
\Delta t=\frac{\lambda }{c}\Delta z=\frac{4\mathbf{\Omega \cdot
S}}{c^{2}} \label{eq:sagnac2}
\end{equation}

Even if his interpretation of these results was entirely in the
framework of the classical (not Lorentz's!) ether theory, Sagnac
was the first scientist who reported an experimental observation
of the effect of rotation in space-time, which, after him, was
named ''Sagnac effect''. It is noteworthy  to notice that the
Sagnac effect has been interpreted as a disproval of the SRT since
the early years of relativity (in particular by Sagnac himself) up
to now (in particular by Selleri\cite{selleri96},\cite{selleri97},
Croca-Selleri\cite {croca99}, Goy-Selleri\cite{goy97},
Vigier\cite{vigier97}, Anastasowski \textit{et
al.}\cite{anastasowski99},
Klauber\cite{klauber},\cite{klauberlibro}). However, this claim is
incorrect. As a matter of fact, the Sagnac effect can be
completely explained in the framework of the SRT: see for instance
Weber\cite{weber97}, Dieks\cite{dieks91}, Anandan\cite{anandan81},
Rizzi-Tartaglia\cite{rizzi98}, Bergia-Guidone \cite{bergia98},
Rodrigues-Sharif\cite{rodrigues01}, Pascual-S\'{a}nchez \textit{et
al.}\cite{pascuallibro}. According to the SRT, eq.
(\ref{eq:sagnac2}) turns out to be just a first order
approximation of the relativistic proper time difference between
counter-propagating light beams. Moreover, in what follows, it
will be apparent that the relativistic interpretation of the
Sagnac effect allows a deeper insight into the very foundations of
the SRT.

Few years before Sagnac, Franz Harres\cite{harres12}, graduate
student in Jena, observed (for the first time but unknowingly) the
Sagnac effect during his experiments on the Fresnel-Fizeau drag of
light. However, only in 1914, Harzer\cite{harzer14} recognized
that the unexpected and inexplicable bias found by Harres was
nothing else than the manifestation of the Sagnac effect.
Moreover, Harres's observations also demonstrated that the Sagnac
fringe shift is unaffected by refraction: in other words, it is
always given by eq. (\ref{eq:sagnac1}), provided that $\lambda $
is interpreted as the light wavelength in a co-moving refractive
medium. So, the Sagnac phase shift depends on the light
wavelength, and not on the velocity of light in the (co-moving)
medium.

If Harres anticipated the Sagnac effect on the experimental ground, Michelson%
\cite{michelson04} anticipated the effect on the theoretical side.
Subsequently, in 1925, Michelson himself and
Gale\cite{michelson25} succeeded in measuring a phase shift,
analogous to the Sagnac's one, caused by the rotation of the
Earth, using a large optical interferometer.

The field of light-optical Sagnac interferometry had a revived
interest after the development of laser (see for instance the
beautiful review paper by Post\cite{post67}, where the previous
experiments are carefully described and their theoretical
implications analyzed). After that, there was an increasing
precision in measurements and a growth of technological
applications, such as inertial navigation\cite{chow85}, where the
''fiber-optical gyro''\cite{vali76} and the ''ring
laser''\cite{stedman97} are used.

\subsection{Universality of the Sagnac Effect}\label{ssec:relativistic}

The experimental data show that: (i) the Sagnac fringe shift
(\ref{eq:sagnac1}) does not depend either on the presence of a
co-moving optical medium or on the group velocity of the
counter-propagating beams; (ii) the Sagnac time difference
(\ref{eq:sagnac2}) does not depend either on the light wavelength
or on the presence of a co-moving optical medium. This is a first
important clue of the universality of the Sagnac effect. However,
the most compelling claim for the universal character of the
Sagnac effect comes from the validity of eq. (\ref {eq:sagnac1})
not  for light beams only, but also for any kind of ''entities''
(such as electromagnetic and acoustic waves, classical particles
and electron Cooper pairs, neutron beams and De Broglie waves and
so on...) travelling in opposite directions along a closed path in
a rotating interferometer, with the same (in absolute value)
velocity with respect to the turntable. This fact is well proved
by experimental texts (see Subsection \ref{ssec:tests}).

Of course the entities take different times for a complete
round-trip, depending on their velocity relative to the turntable;
\textit{but the
difference between these times is always given by eq. (\ref{eq:sagnac2}).}%
{\normalsize \ } So, the amount of the time difference is always
the same, both for matter and light waves, independently of the
physical nature of the interfering beams.

This astounding but experimentally well proved ''universality'' of
the Sagnac effect is quite inexplicable on the basis of the
classical physics, and invokes a geometrical explanation in the
Minkowskian space-time of the SRT.

\subsection{Experimental tests and derivation of the Sagnac Effect}\label{ssec:tests}

The Sagnac effect with matter waves has been verified
experimentally using Cooper pairs\cite{zimmermann65} in 1965,
using neutrons\cite{atwood84} in 1984, using $^{40}Ca$ atoms
beams\cite{riehle91} in 1991 and using electrons, by
Hasselbach-Nicklaus\cite{hasselbach93}, in 1993. The effect of the
terrestrial rotation on neutron phase was demonstrated in 1979 by
Werner et al.\cite{werner79} in a series of famous experiments.

The Sagnac phase shift has been derived, in the full framework of
the SRT, for electromagnetic waves in vacuum (Weber\cite{weber97},
Dieks\cite{dieks91}, Anandan\cite{anandan81},
Rizzi-Tartaglia\cite{rizzi98}, Bergia-Guidone \cite {bergia98},
Rodrigues-Sharif\cite{rodrigues01}). However, a clear and
universally shared derivation for matter waves is not available as
far as we know, or it is at least difficult to find it in the
literature. Indeed, the Sagnac phase shift for matter waves has
been derived,\textit{\ in the first order approximation} with
respect to the velocity of rotation of the interferometer, by many
authors (see Ashby's paper in this book\cite{ashby-libro} and the
paper by Hasselbach-Nicklaus  for discussions and further
references). These derivations are often based on an heterogeneous
mixture of classical kinematics and relativistic dynamics, or non
relativistic quantum mechanics and some relativistic elements.

An example of such derivations is given in a well known paper by Sakurai%
\cite{sakurai80}, on the basis of a formal analogy between the
classical Coriolis force
\begin{equation}
\mathbf{F}_{Cor}=2m_{o}\mathbf{v}\times \mathbf{\Omega \;,}
\label{eq:coriolis1}
\end{equation}
acting on a particle of mass $m_{o}$ moving in a uniformly
rotating frame, and the Lorentz force
\begin{equation}
\mathbf{F}_{Lor}=\frac{e}{c}\mathbf{v}\times \mathbf{B}
\label{eq:lorentz1}
\end{equation}
acting on a particle of charge $e$ moving in a constant magnetic field $%
\mathbf{B}$.

Sakurai considers a beam of charged particles split into two
different paths and then recombined. If $S$ is the surface domain
enclosed by the two paths, the resulting phase difference in the
interference region turns out to be
\begin{equation}
\Delta \Phi =\frac{e}{c\hbar }\int_{S}\mathbf{B}\cdot
\mathrm{d}\mathbf{S} \label{eq:phasemag}
\end{equation}
Therefore, $\Delta \Phi $ is different from zero when a magnetic
field exists inside the domain enclosed by the two paths, even if
the magnetic field felt by the particles along their paths is
zero. This is the well known Aharonov-Bohm\cite{aharonovbohm59}
effect. By formally substituting
\begin{equation}
\frac{e}{c}\mathbf{B}\rightarrow 2m_{o}\mathbf{\Omega }
\label{eq:subsak}
\end{equation}
Sakurai shows that the phase shift (\ref{eq:phasemag}) reduces to
\begin{equation}
\Delta \Phi =\frac{2m_{o}}{\hbar }\int_S \mathbf{\Omega }\cdot
d\mathbf{S} \label{eq:phaseomega}
\end{equation}
If $\mathbf{\Omega }$ is interpreted as the angular velocity
vector of the uniformly rotating turntable and $\mathbf{S}$ as the
vector associated to the area enclosed by the closed path along
which the two counter-propagating material beams travel, then eq.
(\ref{eq:phaseomega}) can be interpreted as the Sagnac phase
shift for the considered counter-propagating beams:\footnote{%
In the case of the Aharonov-Bohm effect, the magnetic field
$\mathbf{B}$ is zero along the trajectories of the particles,
while in  Sakurai's derivation, which we are going to generalize,
the angular velocity $\mathbf{\Omega }$, which is the analogue of
the magnetic field for particles in a rotating frames, is not
null: therefore the analogy with the Aharonov-Bohm effect seems to
be questionable. However, the formal analogy can be easily
recovered when \textit{the flux} of the magnetic field, rather
than the magnetic field itself, is considered: this is just what
we are going to do (see Section \ref {sec:sagnacab}, below).}
\begin{equation}
\Delta \Phi =\frac{2m_{o}}{\hbar }\mathbf{\Omega }\cdot \mathbf{S}
\label{eq:phaseomega1}
\end{equation}
This result has been obtained using non relativistic quantum
mechanics:  the relations between the Aharonov-Bohm effect and the
wave equations are discussed in Subsection \ref{ssec:ab}.

The time difference corresponding to the phase difference (\ref
{eq:phaseomega1}), turns out to be:
\begin{equation}
\Delta t=\frac{\Delta \Phi }{\omega }=\frac{\hbar }{E}\Delta \Phi =\frac{%
\hbar }{mc^{2}}\Delta \Phi =\frac{2m_{o}}{mc^{2}}\mathbf{\Omega
}\cdot \mathbf{S}  \label{eq:deltat}
\end{equation}
Let us point out that eq. (\ref{eq:deltat}) contains,
inconsistently but unavoidably, some relativistic elements ($\hbar
\omega =E=mc^{2}$). Of course in the first order approximation,
i.e. when the relativistic mass $m$
coincides with the rest mass $m_{o}$, eq. (\ref{eq:deltat}) reduces to eq. (%
\ref{eq:sagnac2}); that is, as we stressed before, a first order
approximation of the relativistic time difference associated to
the Sagnac effect.\footnote{%
Formulas (\ref{eq:sagnac2}) and (\ref{eq:deltat}) differ by a
factor 2: this depends on the fact that in eq. (\ref{eq:sagnac2})
we considered the complete round-trip of the beams, while in this
section we refer to a situation in which the emission point and
the interference point are diametrically opposed.\label{fn:fact2}}

This  simple and beautiful procedure will be generalized and
extended to a fully relativistic context in Sec.
\ref{sec:sagnacab}.

\section{Direct derivation: Sagnac effect for material and light particles}\label{sec:direct}

\subsection{Direct derivation}\label{ssec:ddirect}

In this section we are going to give a description of light or
matter beams counter-propagating in a rotating interferometer,
based on  relativistic kinematics; here and henceforth, we shall
refer to both light and matter beams by calling them simply
''beams''. Indeed, it is our aim to show that, under suitable
conditions, the Sagnac time difference does not depend on the very
physical nature of the interfering beams.\newline

The beams are constrained to travel a circular path along the rim
of the rotating disk, with constant angular velocity, in opposite
directions. Let us suppose that a beam source and an
interferometric detector are lodged on a point $\Sigma $ of the
rim of the disk. Let $K$ be the central inertial frame,
parameterized by an adapted (see Appendix A.\theCiSei) set of
cylindrical coordinates $\left\{ x^{\mu }\right\}
~=~\left( t,r,\vartheta ,z\right) $, with line element given by\footnote{%
The signature is (-1,1,1,1), Greek indices run from 0 to 3, while
Latin indices run from 1 to 3.}

\begin{equation}
\mathrm{d}s^{2}=g_{\mu \nu }dx^{\mu }dx^{\nu }=-c^{2}\mathrm{d}t^{2}+\mathrm{%
d}r^{2}+r^{2}\mathrm{d}\vartheta ^{2}+\mathrm{d}z^{2}\mathrm{\ }
\label{eq:metricapiatta}
\end{equation}
In particular, if we confine ourselves to a the disk ($z=const$),
the metric turns out to be
\begin{equation}
ds^2=-c^2dt^2+dr^2+r^2d\vartheta^2  \label{eq:metrica}
\end{equation}

With respect to $K$, the disk (whose radius is $R$) rotates with
angular velocity $\Omega $, and the world-line $\gamma _{\Sigma }$
of $\Sigma $ is
\begin{equation}
\gamma _{\Sigma }\equiv \left\{
\begin{array}{rcl}
x^{0} & = & ct \\
x^{1} & = & r=R \\
x^{2} & = & \vartheta =\Omega t
\end{array}
\right.  \label{eq:gammasigma0}
\end{equation}
or, eliminating $t$
\begin{equation}
\gamma _{\Sigma }\equiv \left\{
\begin{array}{rcl}
x^{0} & = & \frac{c}{\Omega }\vartheta \\
x^{1} & = & R \\
x^{2} & = & \vartheta
\end{array}
\right.  \label{eq:gammasigma}
\end{equation}
The proper time read by a clock at rest in $\Sigma $ is given by
\begin{equation}
\tau =\frac{1}{c}\int_{\gamma _{\Sigma
}}|ds|=\frac{1}{c}\int_{\gamma _{\Sigma
}}\sqrt{c^{2}dt^{2}-R^{2}d\vartheta ^{2}}=\frac{1}{\Omega }\sqrt{1-\beta ^{2}}%
\int_{\gamma _{\Sigma }}d\vartheta  \label{eq:dtau}
\end{equation}
\begin{figure}[th]
\begin{center}
\includegraphics[width=8cm,height=8cm]{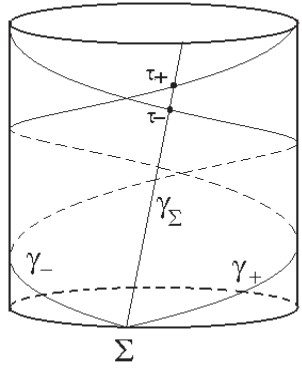}
\caption{\small The world-line of $\Sigma$ (a point on the disk
where a beam source and interferometric detector  are lodged) is
$\gamma_\Sigma$;  $\gamma_+$ and $\gamma_-$ are the world-lines of
the co-propagating (+) and counter-propagating (-) beams. The
first intersection of $\gamma _{+}$ ($\gamma _{-}$) with $\gamma
_{\Sigma }$  takes place at the  time $\tau_+$ ($\tau_-$), as
measured by a clock at rest in $\Sigma$.} \label{fig:helix}
\end{center}
\end{figure}
\normalsize

The world-lines of the co-propagating (+) and counter-propagating
(-) beams emitted by the source at time $t=0$ (when $\vartheta
=0$) are, respectively:
\begin{equation}
\gamma _{+}\equiv \left\{
\begin{array}{rcl}
x^{0} & = & \frac{c}{\omega _{+}}\vartheta \\
x^{1} & = & R \\
x^{2} & = & \vartheta
\end{array}
\right.  \label{eq:gammapiu}
\end{equation}
\begin{equation}
\gamma _{-}\equiv \left\{
\begin{array}{rcl}
x^{0} & = & \frac{c}{\omega _{-}}\vartheta \\
x^{1} & = & R \\
x^{2} & = & \vartheta
\end{array}
\right.  \label{eq:gammameno}
\end{equation}
where $\omega _{+},\omega _{-}$ are their angular velocities, as
seen in the
central inertial frame.\footnote{%
Notice that $\omega _{-}$ is positive if $|\omega _{-}^{\prime
}|<\Omega $, null if $|\omega _{-}^{\prime }|=\Omega $, and
negative if $|\omega _{-}^{\prime }|>\Omega $; see
eq.(\ref{eq:betaunobeta}) below.} The first intersection of
$\gamma _{+}$ ($\gamma _{-}$) with $\gamma _{\Sigma }$ is the
event ''absorption of the co-propagating (counter-propagating)
beam after a complete round trip'' (see figure \ref{fig:helix}).

This event takes place when
\begin{equation}
\frac{1}{\Omega }\vartheta _{\pm }=\frac{1}{\omega _{\pm
}}(\vartheta _{\pm }\pm 2\pi )  \label{eq:round}
\end{equation}
where the $+$ ($-$) sign holds for the co-propagating
(counter-propagating) beam. The solution of eq. (\ref{eq:round})
is:
\begin{equation}
\vartheta _{\pm }=\pm \frac{2\pi \Omega }{\omega _{\pm }-\Omega }
\label{eq:thetapiu}
\end{equation}
If we introduce the dimensionless velocities $\beta =\Omega R/c$,
$\beta _{\pm }=\omega _{\pm }R/c$, the $\vartheta $-coordinate of
the absorption event can be written as follows:
\begin{equation}
\vartheta _{\pm }=\pm \frac{2\pi \beta }{\beta _{\pm }-\beta }
\label{eq:thetapiumeno}
\end{equation}
Taking into account eq. (\ref{eq:dtau}), the proper time elapsed
between the emission and the absorption of the co-propagating
(counter-propagating) beam, read by a clock at rest in $\Sigma $,
is given by
\begin{equation}
\tau _{\pm }=\ \ \pm \frac{2\pi \beta }{\Omega }\frac{\sqrt{1-\beta ^{2}}}{%
\beta _{\pm }-\beta }  \label{eq:taupiumeno}
\end{equation}
and the proper time difference $\Delta \tau \equiv \tau _{+}-\tau
_{-}$ turns out to be
\begin{equation}
\Delta \tau =\frac{2\pi \beta }{\Omega }\sqrt{1-\beta
^{2}}\frac{\beta _{-}-2\beta +\beta _{+}}{(\beta _{+}-\beta
)(\beta _{-}-\beta )} \label{eq:deltatau11}
\end{equation}

Without specifying any  conditions, the proper time difference
(\ref {eq:deltatau11}) appears to depend upon $\beta ,\beta
_{+},\beta _{-}$: this means that it does depend, in general, both
on the velocity of rotation of the disk and on the velocities of
the beams. Let $\beta _{\pm }^{\prime }$ be the velocities of the
beams as measured in any Minkowski inertial frame, locally
co-moving with the rim of the disk, or briefly speaking in any
locally co-moving inertial frame (LCIF). Provided that each LCIF
is Einstein-synchronized (see Subsection \ref{ssec:synchro}
below), the Lorentz law of velocity addition gives the following
relations between $\beta _{\pm }^{\prime }$ and $\beta _{\pm }$:
\begin{equation}
\beta _{\pm }=\frac{\beta _{\pm }^{\prime }+\beta }{1+\beta _{\pm
}^{\prime }\beta }  \label{eq:betaunobeta}
\end{equation}
By substituting (\ref{eq:betaunobeta}) in (\ref{eq:deltatau11}) we
easily obtain
\begin{equation}
\Delta \tau =\frac{4\pi \beta ^{2}}{\Omega }\frac{1}{\sqrt{1-\beta ^{2}}}+%
\frac{2\pi \beta }{\Omega }\frac{1}{\sqrt{1-\beta ^{2}}}\left( \frac{1}{%
\beta _{+}^{\prime }}+\frac{1}{\beta _{-}^{\prime }}\right)
\label{eq:deltatau12}
\end{equation}

Now, let us impose the condition ''equal relative velocity in
opposite directions'':
\begin{equation}
\beta _{+}^{\prime }=-\beta _{-}^{\prime }  \label{eq:lec}
\end{equation}

This condition means that the beams are required to have the same
velocity
(in absolute value) in every LCIF,\footnote{%
Or, differently speaking, with respect to any observer at rest in
the ''relative space'' (see below) along the rim of the platform.}
provided that every LICF is Einstein-synchronized. If condition
(\ref{eq:lec}) is imposed, the proper time difference
(\ref{eq:deltatau12}) reduces to
\begin{equation}
\Delta \tau =\frac{4\pi \beta ^{2}}{\Omega }\frac{1}{\sqrt{1-\beta ^{2}}}=%
\frac{4\pi R^{2}\Omega }{c^{2}}\left( 1-\frac{\Omega ^{2}R^{2}}{c^{2}}%
\right) ^{-1/2}  \label{eq:deltatausagnac}
\end{equation}
which is the relativistic Sagnac time difference.

A very relevant conclusion follows. According to eq.
(\ref{eq:taupiumeno}), the beams take different times - as
measured by the clock at rest on the starting-ending point $\Sigma
$ on the platform - for a complete round trip, depending on their
velocities $\beta _{\pm }^{\prime }$ relative to the
turnable. However, when condition (\ref{eq:lec}) is imposed, the difference $%
\Delta \tau $ between these times does depend \textit{only} on the
angular velocity $\Omega $ of the disk, and it does not depend on
the velocities of propagation of the beams with respect the
turnable.

This is a very general result, which has been obtained on the
ground of a purely kinematical approach. The Sagnac time
difference (\ref{eq:deltatausagnac}) applies to any couple of
(physical or even mathematical) entities, as long as a velocity
with respect the turnable can be consistently defined. In
particular, this result applies as well to photons (for which
$|\beta _{\pm }^{\prime }|=1$) and to any kind of classical or
quantum particles under the given conditions (or
electromagnetic/acustic waves in presence of an homogeneous
co-moving medium). \footnote{Provided that a group velocity can be
defined.} This fact highlights, in a clear and straightforward
way, the universality of the Sagnac effect.

More in particular, the Sagnac time difference
(\ref{eq:deltatausagnac}) also applies to the Fourier components
of the wave packets associated to a couple of matter beams
counter-propagating (with the same relative velocity) along the
rim.\footnote{Of course only matter beams are physical entities,
while Fourier components are just mathematical entities, to which
no energy transport is associated.} This remark is important  to
study the interferometric detectability of the Sagnac effect  (see
Subsection \ref{ssec:detsagnac} below).

\subsection{The Sagnac effect as an empirical evidence of the SRT}\label{ssec:empsagnac}

As we said in the Introduction, the Sagnac effect for
electromagnetic waves in vacuum was first interpreted by Sagnac
himself as an experimental evidence of the physical existence of
the classical (non relativistic) ether, and an experimental
disproval of the SRT. Sagnac's interpretation can be
easily understood, on the basis of the relativistic eq. (\ref{eq:deltatau11}%
), as a casual consequence of a well known kinematical feature of
light propagation through the ether. Indeed, the light velocity
with respect to the ether (at rest in the central IF) must be $c$
in both directions; as a consequence, if we set $\beta _{\pm }=\pm
1$ in eq. (\ref{eq:deltatau11}), the
proper time difference $\Delta \tau $ reduces to eq. (\ref{eq:deltatausagnac}%
); the latter, in turn, reduces, at first order approximation, to
the time difference given in eq. (\ref{eq:sagnac2}), which was
actually predicted and experimentally tested by Sagnac.

However, any non relativistic explanation completely fails for
subluminally travelling entities (such as matter waves, sound
waves, electromagnetic waves in an homogeneous co-moving medium,
and so on). In fact, for subluminally travelling entities the
vital condition is: ''equal relative velocity in opposite
directions''. If this condition is expressed by eq. (\ref{eq:lec})
(which explicitly requires the local Einstein synchronization) the
Sagnac proper time difference (\ref{eq:deltatausagnac}) arises, as
we carefully showed before.

On the contrary, if the condition ''equal relative velocity in
opposite directions'' is expressed by an analogous relation, in
which the  local Einstein synchronization is replaced by a
synchronization borrowed from the global synchronization of the
central IF (see Subsection \ref{ssec:synchro}), no time difference
arises. Let us prove this claim.\\

Let $\varphi _{\pm }$ and $\vartheta _{\pm }^{r}$ be the azimuthal
coordinates of the co-rotating/counter-rotating entity with
respect to the central IF and the LCIF, respectively. These
coordinates are related by the transformation
\begin{equation}
\varphi _{\pm }=\Omega t+\vartheta^r_{\pm }  \label{eq:fi+-}
\end{equation}

Derivation of eq. (\ref{eq:fi+-}) with respect to the central
inertial time $t$ gives
\begin{equation}
\omega _{\pm }=\Omega +\omega _{\pm }^{r}  \label{eq:omega+-}
\end{equation}
where $\omega _{\pm }\equiv d\varphi _{\pm }/dt$ is the angular
velocity relative to the central IF, and $\omega _{\pm }^{r}\equiv
d\vartheta _{\pm }^{r}/dt$ is the angular velocity relative to the
LCIF - provided that it is synchronized by means of the central
inertial time $t$.

Eq. (\ref{eq:omega+-}), multiplied by R/c, takes the dimensionless
form
\begin{equation}
\beta _{\pm }=\beta +\beta _{\pm }^{r}  \label{eq:beta+-}
\end{equation}
which  replaces eq. (\ref{eq:betaunobeta}). Let us stress that both eqs. (%
\ref{eq:betaunobeta}) and (\ref{eq:beta+-}) are correct: the
former refers to the local Einstein synchronization, the latter to
the local synchronization according to the simultaneity criterium
of the central IF.

Introducing eq. (\ref{eq:beta+-}) into eq. (\ref{eq:deltatau11}),
the proper time difference $\Delta \tau $ reduces to
\begin{equation}
\Delta \tau =\frac{2\pi \beta }{\Omega }\sqrt{1-\beta
^{2}}\frac{(\beta +\beta _{-}^{r})-2\beta +(\beta +\beta
_{+}^{r})}{\beta _{-}^{r}\beta _{+}^{r}}=\frac{2\pi \beta }{\Omega
}\sqrt{1-\beta ^{2}}\frac{\beta _{-}^{r}+\beta _{+}^{r}}{\beta
_{-}^{r}\beta _{+}^{r}}  \label{eq:bla}
\end{equation}

If the vital condition ''equal relative velocity in opposite
directions'' is expressed by eq.
\begin{equation}
\beta _{+}^{r}=-\beta _{-}^{r}  \label{eq:lecr}
\end{equation}
instead of eq. (\ref{eq:lec}), it is plain from eq. (\ref{eq:bla})
that no time difference arises: $\Delta \tau =0$ (Q.E.D.)

This  calculation shows that the choice of the local Einstein
synchronization is crucial even in non-relativistic motion.
Indeed,  the choice of the classical eq. (\ref{eq:beta+-}),
instead of the relativistic eq. (\ref{eq:betaunobeta}), could be
naively presumed as a reasonable approximation in non-relativistic
motion: however, this choice simply cancels the effect!

This shows that, according to a very appropriate remark by Dieks
and Nienhuis \cite{dieks90}, the observed Sagnac effect is an
experimental evidence of the SRT at first order approximation with
respect to $\beta=\Omega R/c$, and an experimental disproval of
the classical
(non relativistic) ether. \\

\textbf{Remark }It could be always possible to substitute eq.
(\ref{eq:lec}) with an alternative suitable condition, so that eq.
(\ref{eq:bla}) turns out to be equal to eq.
(\ref{eq:deltatausagnac}). Such a condition is:
\begin{equation}
\beta _{-}^{r}=-\beta _{+}^{r}\frac{1-\beta ^{2}}{1-2\beta \beta
_{+}^{r}-\beta ^{2}} \label{eq:lecr_iner}
\end{equation}
But, of course, this is an extremely ''ad hoc'' condition, which
translates  the simple and expressive condition (\ref{eq:lec}): it
is clear that the physical interpretation  of (\ref{eq:lecr_iner})
is not as evident  as that of (\ref{eq:lec}).

\subsection{Interferometric detectability of the Sagnac effect}\label{ssec:detsagnac}

With regard to the interferometric detection of the Sagnac effect,
the crucial point is the following one. Consider the Fourier
components of the wave packets associated to a couple of matter
beams counter-propagating, with the same group velocity, along the
rim. Despite the lack of a direct physical meaning and energy
transfer, the phase velocity of these Fourier components complies
with the Lorentz law of velocity addition (\ref{eq:betaunobeta} ),
and is the same for both the co-rotating and counter-rotating
Fourier components. As a consequence, the Sagnac time difference
(\ref {eq:deltatausagnac}) also applies to any couple of Fourier
components with the same phase velocity.

Moreover, the interferometric detection of the Sagnac effect
requires that the wave packet associated to the matter beam should
be sharp enough in the frequency space, to allow the appearance,
in the interferometric region, of an observable fringe
shift.\footnote{That is, the Fourier components of the  wave
packet should have slightly different wavelengths.} It may be
worth recalling that:

(i) the observable fringe shift $\Delta z$ depends on  the phase
velocity of the Fourier components of the wave packet;

(ii) with respect to an Einstein-synchronized LCIF, the velocity
of every Fourier component of the wave packet associated to the
matter beam, moving
with the velocity (in absolute value) $v\equiv c|\beta _{\pm }^{\prime }|$,%
\textbf{\ }is given by the De Broglie expression $v_{f}=c^{2}/v$.

The consequent Sagnac phase shift, due to the relativistic time difference (%
\ref{eq:deltatausagnac}), is
\begin{equation}
\Delta \Phi =2\pi \Delta z=2\pi \left( \frac{v_{f}}{\lambda
}\Delta \tau \right) =\frac{8\pi ^{2}R^{2}\Omega }{\lambda
v}\left( 1-\frac{\Omega ^{2}R^{2}}{c^{2}}\right) ^{-1/2}
\label{eq:phaseshift}
\end{equation}

\subsection{Comparing to some results found in the literature}\label{ssec:comparlit}

As we mentioned, the Sagnac time delay for matter beams has been
derived by many authors in many different ways, but almost always
in the first order approximation. However, digging into the
literature, we eventually found, between the preliminary and the
final version of this paper, a couple of derivations of the Sagnac
effect for matter beams in full SRT. In this section we are going
to compare these approaches to ours.

\subsubsection{Dieks-Nienhuis's approach}\label{sssec:dieksnien}

Dieks and Nienhuis\cite{dieks90} move from the standard Lorentz
transformation from the LCIF to the central IF. In order to be as
clear and self-consistent as possible, we shall translate
everything into our notations.

Consider two near events, happening along the rim, belonging to
the world-line of the co-rotating/counter-rotating beam; let $dx$,
d$\tau $ be the space and time separation between these events, as
measured in an Einstein-synchronized LCIF. Then the corresponding
time separation $dt$, as measured in the central IF, is given by
the usual Lorentz transformation
\begin{equation}
dt=\left( d\tau \pm \frac{\Omega Rdx}{c^{2}}\right) \left(
1-\frac{\Omega ^{2}R^{2}}{c^{2}}\right) ^{-1/2} \label{eq:lorentz}
\end{equation}
where the $+$ and $-$ hold for the co-rotating and
counter-rotating beams, respectively. Gluing together all  LCIFs,
at the end of the  round trip we have
\begin{equation}
\left( 1-\frac{\Omega ^{2}R^{2}}{c^{2}}\right) ^{1/2}t_{\pm }=\tau
(\gamma_\pm) \pm \frac{\Omega R}{c^{2}}\int_{D}dx
\label{eq:dieks1}
\end{equation}
where $D$ is the length of the rim of the platform, as seen on the
platform itself.

The difference between the   equations for the co-rotating and
counter-rotating beams is:
\begin{equation}
\left( 1-\frac{\Omega ^{2}R^{2}}{c^{2}}\right)
^{1/2}(t_{+}-t_{-})=\tau (\gamma_+)-\tau (\gamma_-)+\frac{2\Omega
R}{c^{2}}\int_{D}dx \label{eq:dieks2}
\end{equation}

In the derivation given by  Dieks and Nienhuis, two hypotheses
play a vital role in order to get the correct conclusion: (i) the
integration domain $D$ is $2\pi R\left( 1-\frac{\Omega
^{2}R^{2}}{c^{2}}\right) ^{-1/2};$ (ii) $\tau
(\gamma_+)=\tau (\gamma_-)=\left( \frac{2\pi R}{v^{\prime }}\right) \left( 1-\frac{%
\Omega ^{2}R^{2}}{c^{2}}\right) ^{-1/2},$ where $v'$ is the
relative velocity (in absolute value) of both the co-rotating and
the counter-rotating beam (as a consequence, $\tau (\gamma_+)-\tau
(\gamma_-)=0$).

Of course, the two hypotheses are correct, but both deserve
further remarks. First of all, we could say that hypothesis (i) is
unnecessary: indeed, our derivation of the Sagnac effect does not
depend on the hyperbolic features of the space geometry of the
disk (see Subsection \ref{ssec:relspace} and Appendix
A.\theCiQuattordici). On the other hand, this hypothesis
establishes a very interesting link between the (observable)
Sagnac time delay and the (unobservable) lengthening of the rim of
the rotating platform, which, in turn, depends on the peculiar
space geometry of the disk. This is a challenge to those authors
(see for instance \cite{klauberlibro},\cite{tartaglia-libro}) who
try to conciliate the Sagnac time delay with the Euclidean space
geometry of the disk.

\subsubsection{Malykin's approach}\label{sssec:maly}

The long review paper by Malykin\cite{malykin2000} (almost 300
references!) starts with the remark that the Sagnac effect for
matter beams ''is explained in several totally different ways'',
but - strangely enough - most of them give ''correct results
despite their obvious incorretness''. After a wide overview of
these ''incorrect explanations'', the author provides his attempt
of solution on the basis of the relativistic law of velocity
addition. This approach is the only one that turns out to be
similar to ours; also the problem of the interferometric detection
of the Sagnac effect is taken into account. Of course, this leads
to the issue of the kinematical behavior of both the phase
velocity and the group velocity. Here some misinterpretations and
confusions arise. In particular, the author tries to show that
''both the group velocity and the phase velocity have identical
translational properties during the transition from the frame of
reference K [the central IF] to the frame of reference K' [the
LCIF]''.

Actually, the  group velocity behaves contrary to the phase
velocity with respect to the (local) Lorentz transformation group.
However, this has no consequences on the interferometric detection
of the Sagnac effect, because the only relevant requirement is
that, with respect to any Einstein-synchronized LCIF: (i) the
group velocities should be the same; (ii) the phase velocities of
the Fourier components should be the same. This is exactly what
happens: the transformation properties have no role. Anyway, we
agree with Malykin when he says that ''the Sagnac effect
constitutes a kinematical effect of SRT; (...) all explanations of
the Sagnac effect are incorrect except the relativistic one''. We
agree also with his final remark: the existence of a large number
of incorrect explanations which give correct results (at least in
first approximation) depends on the fact that the Sagnac effect is
a first-order effect in $v/c$.

\subsubsection{Anderson, Stedman and Bilger's approach}\label{sssec:andstedbil}

It is interesting to compare our results to this approach, even
though Anderson, Stedman and Bilger's
results\cite{stedman97},\cite{anderson94} are confined to a first
order approximation. Indeed, they find, at first order
approximation, the following time difference:
\begin{equation}
\Delta t=\frac{4\mathbf{\Omega \cdot S}}{v^{2}} \label{eq:stedman}
\end{equation}
and the following phase shift:
\begin{equation}
\Delta \Phi =\frac{8\pi \mathbf{\Omega \cdot S}}{\lambda v}
\label{eq:stedmanfase}
\end{equation}
where $v$ is the ''undragged'' velocity of the beams. Of course,
the time difference (\ref{eq:stedman}) is not in
agreement\footnote{Except for light beams in vacuum.} with the
first order approximation (with respect to $\beta =\Omega R/c$) of
eq. (\ref {eq:deltatausagnac}). However, it is consistent with the
first order approximation of eq. (\ref{eq:deltatau11}) provided
that $\beta _{+}=-\beta _{-}\equiv v/c$: this represents a
completely different physical situation,\footnote{Except for light
beams in vacuum.} in which the two beams are injected into the
rotating platform (tangentially to the rim) in opposite directions
\textit{with the same velocity with respect to the central
inertial frame. }

In addition, the phase shift (\ref{eq:stedmanfase}), which is the
only observable quantity through an interferometric device, is not
in agreement with the physical situation considered by these
authors. However, it is worthwhile to notice that, oddly enough,
it perfectly agrees with the first order approximation of eq.
(\ref {eq:phaseshift}).\footnote{If condition (\ref{eq:lec}) is
imposed, eq. (\ref{eq:stedman}) is wrong and eq.
(\ref{eq:stedmanfase}) is right. However, both of them are right
for light beams in vacuum.}

\subsubsection{Ashby's approach}\label{ssec:ashby}

The best approach  we know, at first order approximation, is the
one suggested in this book by Ashby \cite{ashby-libro}. The
peculiarity of this approach is that it is independent of the
shape of the loop: this is a very important feature if one has to
deal with the Global Positioning System (GPS). In particular,
Ashby  shows that the Sagnac time delay depends only on the area
swept out by the electromagnetic pulse, as it travels from the GPS
transmitter to the receiver, projected onto the terrestrial
equatorial plane. A great care is devoted to synchronization
problems: this issue is in complete agreement with our fully
relativistic approach (see Subsection \ref{ssec:synchro} below).

\subsection{Synchronization in a LCIF: a free choice}\label{ssec:synchro}

As pointed out by Rizzi-Serafini\cite{rizzi-serafini-libro}, in a
local or global inertial frame (IF) the synchronization is not
''given by God'', as often both relativistic and anti-relativistic
authors assume, but it can be arbitrarily chosen within the
\textit{synchronization gauge}
\begin{equation}
\left\{
\begin{array}{l}
t^{\prime }=t^{\prime }\,(\,t,\,x^{1},x^{2},x^{3}) \\
x_{i}^{\prime }=x_{i}
\end{array}
\right.   \label{eq:synchr}
\end{equation}
The synchronization gauge (\ref{eq:synchr}) is a subset of the \textit{%
Cattaneo gauge} (\ref{eq:gauge_trans}) (see Appendix
A.\theCiQuattro), which is the set of all  possible
parameterizations of the given physical inertial frame (IF). In
eq. (\ref
{eq:synchr}) the coordinates ($t,x_{i}$) are Einstein coordinates, and ($%
t^{\prime },x_{i}^{\prime }$) are re-synchronized coordinates of
the IF under consideration. Of course, the IF turns out to be
optically isotropic if and only if it is parameterized by Einstein
coordinates ($t,x_{i}$). Then the following question arises: if
the parameterization (in particular the synchronization) of a LCIF
is a matter of choice, which is the most profitable choice in
order to describe the Sagnac effect?

Since the synchronization gauge (\ref{eq:synchr}) is too general
for a clear and useful discussion, it is advantageous to introduce
a suitable subset of the synchronization gauge that allows a
clearer discussion. Such a sub-gauge actually exists; it has been
introduced by Selleri\cite{selleri96},\cite{selleri97}. Let us
briefly summarize the so-called "Selleri gauge".

Let $K$ be a ''formally privileged'' IF, in which an isotropic
synchronization (that is Einstein synchronization) is assumed
\textit{by
stipulation; }and let $S$ be  a (generally anisotropic) IF moving along the $%
x_{1}^{\prime }=x_{1}$ axis with dimensionless velocity $\beta $
with respect to $K$. The \textit{Selleri gauge }is defined by
\begin{equation}
\left\{
\begin{array}{l}
t^{\prime }=t+\frac{\Gamma (\beta )}{c}x \\
x_{i}^{\prime }=x_{i}
\end{array}
\right.   \label{eq:selleri-gauge}
\end{equation}
where the unprimed coordinates refer to the Einstein-synchronized
IF $K$, the primed coordinates refer to the
arbitrarily-synchronized IF $S$ and $\Gamma (\beta )$ is an
arbitrary function of $\beta $. It is convenient to write this
function as follows:
\begin{equation}
\Gamma (\beta )\equiv \beta +e_{1}(\beta )c\gamma ^{-1}
\label{eq:gammaM}
\end{equation}
The  function $e_{1}(\beta )$ is  Selleri's ''synchronization
parameter'', which, in principle, can be arbitrarily chosen. Any
choice of the function $e_{1}(\beta )$ is a choice of the
synchronization in the IF under consideration; in principle, the
synchronization can be freely chosen inside the Selleri gauge
(\ref{eq:selleri-gauge}). In particular, the synchrony choice
$e_{1}(\beta )\doteq -\beta \gamma /c$ (i.e. $\Gamma (\beta
)\doteq 0$) gives the standard Einstein synchronization, which is
''relative'' (that is, frame-dependent); whereas the synchrony choice $%
e_{1}(\beta )\doteq 0$ gives the Selleri synchronization, which is
''absolute'' (that is, frame-independent).

The term ''absolute'' sounds rather eccentric in a relativistic
framework, but it simply means that the\ Selleri simultaneity
hypersurfaces $t^{\prime }= const$ (contrary to the  Einstein
simultaneity hypersurfaces $t= const$) define a frame-invariant
foliation of space-time - which is nothing but the Einstein
foliation of the particular IF $K$ assumed (by stipulation, once
and for all) as optically isotropic for any choice of the
synchronization parameter.

According to Selleri, the synchronization is a matter of
convention in the case of translation, but not in the case of
rotation: when rotation is taken into account, the synchronization
parameter $e_{1}$ is forced to take the value zero. On the
contrary, as it is shown in \cite{rizzi-serafini-libro}, the
choice of $e_{1}$ is not compelled by any empiric evidence: that
is,
also when rotation is taken into account, no physical effect can discriminate%
{\ } Selleri's synchrony choice $e_{1}(\beta )\doteq 0$ from the
Einstein's synchrony choice $e_{1}(\beta )\doteq -\beta \gamma
/c${. }

{Therefore, we have the opportunity of taking a very pragmatic
view: both Selleri ''absolute'' synchronization and Einstein
relative synchronization can be used, }depending on the aims and
circumstances. In particular, (i) if
we look for a global synchronization on the rotating platform, Selleri {%
''absolute'' synchronization is required; (ii) if we look for a
plain kinematical relationship between local velocities, Einstein
synchronization is required in any LCIF. }

Let us outline the advantages of the local Einstein
synchronization on a rotating platform. First, let us
recall\cite{rizzi-serafini-libro} that the local isotropy or
anisotropy of the velocity of light in a LCIF is not a fact, with
a well defined ontological meaning, but a convention which depends
on the synchronization chosen in the LCIF. Of course the velocity
of light has the invariant value $c$ in every LCIF, both in
co-rotating and counter-rotating direction, if and only if the
LCIFs are Einstein-synchronized. We are aware that this statement
sounds irrelevant or arbitrary to some authors\cite{klauber},
\cite{selleri96}, \cite{selleri97}, \cite{peres}, so we  try to
suggest a more significant one. As we showed in Subsection.
\ref{ssec:ddirect}, the Sagnac time difference
(\ref{eq:deltatausagnac}) holds for two beams travelling along the
rim in opposite directions with the same velocity with respect the
turnable. This is a plain and meaningful condition: but it must be
stressed that \textit{this condition requires that every LCIF
should be Einstein-synchronized}{.}\ Of course this condition
could be translated also into  Selleri absolute synchronization,
but it would result in a very artificial and convoluted
requirement, namely eq. (\ref{eq:lecr_iner}). Only Einstein
synchronization allows the clear and meaningful requirement:
''equal relative velocity in opposite
directions''.\footnote{Formally expressed by condition
(\ref{eq:lec}).}

\section{The Sagnac effect from an analogy with the Aharonov-Bohm
effect}\label{sec:sagnacab}

In this section we shall give another derivation of the Sagnac
effect for relativistic material beams counter-propagating on a
rotating disk\cite{rizzi03}. This derivation is based on a
(formal) analogy with the Aharonov-Bohm effect which has been
outlined by Sakurai\cite{sakurai80}. However, Sakurai's approach,
which rests upon the use of relativistic and Newtonian elements,
gives only the first order  approximation of Sagnac time
difference (\ref{eq:deltatausagnac}). We want to show that, using
Cattaneo's splitting techniques, it is possible to state the
analogy between the Aharonov-Bohm effect and the Sagnac effect in
a fully relativistic context, getting rid of the Newtonian
elements  and recovering the relativistic self-consistency of the
derivation. On equal footing, our approach allows us to obtain the
Sagnac time difference in full theory and not in its first order
approximation, to which Sakurai and other authors confined
themselves, exploiting the analogy with the Aharonov-Bohm effect.

According to us, our derivation  highlights  the common geometric
nature of the two effects, which is the basis of their
far-reaching analogy.\\

\begin{figure}[ht]
\begin{center}
\includegraphics[width=8cm,height=8cm]{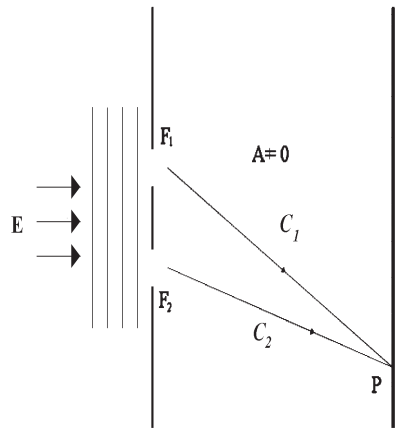}
\caption{\small A single coherent charged beam, originating in
$E$, is split into two parts (passing through the two slits $F_1$
and $F_2$) that propagate, respectively,  along the paths $C_1$
and $C_2$ (in the figure these paths are represented,
respectively, by  $EF_1P$ and $EF_2P$).  The beams travel in a
region where a vector potential $\mathbf{A}$ is present. In $P$,
the beams interfere and an additional phase shift is provoked by
the magnetic field.} \label{fig:ab1}
\end{center}
\end{figure}
\normalsize

\subsection{The Aharonov-Bohm effect}\label{ssec:ab}

\begin{figure}[ht]
\begin{center}
\includegraphics[width=8cm,height=8cm]{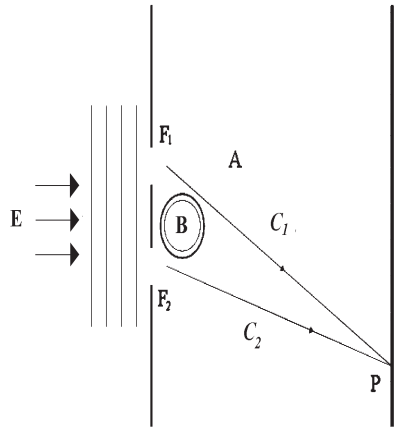}
\caption{\small A single coherent charged beam, originating in
$E$, is split into two parts (passing through the two slits $F_1$
and $F_2$) that propagate, respectively,  along the paths $C_1$
and $C_2$ (in the figure these paths are represented,
respectively, by  $EF_1P$ and $EF_2P$). Between the paths a
solenoid is present; the magnetic field $\mathbf{B}$ is entirely
contained inside the solenoid, while outside there is a constant
vector potential $\mathbf{A}$. In $P$, the beams interfere and an
additional phase shift, provoked by the magnetic field confined
inside the solenoid, is observed.} \label{fig:ab2}
\end{center}
\end{figure}

Let us start by briefly describing the Aharonov-Bohm effect.
Consider the two slits experiment (figure \ref{fig:ab1}) and
imagine that a single coherent charged beam is split into two
parts, which travel in a region where only a magnetic field is
present, described by the 3-vector potential ${\bf A}$; then the
beams are recombined to observe the interference pattern. The
phase  of the two wave functions, at each point of the pattern, is
modified, with respect to the case of free propagation (${\bf
A}~=~0$), by the magnetic potential. The magnetic
potential-induced phase shift has the form\cite{aharonovbohm59}
\begin{equation}
\Delta \Phi=\frac{e}{c\hbar }\oint_{C}{\bf A}\cdot {\rm d}{\bf r}=\frac{e}{%
c\hbar }\int_{S}{\bf B}\cdot {\rm d}{\bf S}  \label{eq:ab}
\end{equation}
where $C$ is the oriented closed curve, obtained as the sum of the
oriented paths $C_{1}$ and $C_{2}$ relative to each component of
the beam (in the physical space, see figure \ref{fig:ab1}). Eq.
(\ref{eq:ab}) expresses (by means of  Stoke's Theorem) the phase
difference in terms of the flux of the magnetic field across the
surface $S$ enclosed by the curve $C$. Aharonov and
Bohm\cite{aharonovbohm59} applied this result to the situation in
which the two  split beams pass one on each side of a solenoid
inserted between the paths (see figure \ref{fig:ab2}). Thus, even
if the magnetic field ${\bf B}$ is totally contained within the
solenoid and the beams pass through a ${\bf B}~=~0$ region, a
resulting phase shift appears, since a non null magnetic flux is
associated to every closed path which encloses the
solenoid.\\

Tourrenc\cite{tourrenc77} showed that no explicit wave equation is
demanded to describe the Aharonov-Bohm effect, since its
interpretation is a pure geometric one: in fact eq. (\ref{eq:ab})
is independent of the  nature of the interfering charged beams,
which can be spinorial, vectorial or tensorial. So, if we deal
with relativistic charged beams, their propagation is described by
a relativistic wave equation, such as the Dirac equation or the
Klein-Gordon equation, depending on the nature of the beams
themselves . From a physical point of view,  spin has no influence
on the Aharonov-Bohm effect because there is no coupling with the
magnetic field which is confined inside the solenoid. Moreover, if
the magnetic field is null, the Dirac equation is equivalent to
the Klein-Gordon equation, and this is the case of a constant
potential. As far as we are concerned,  since in what follows we
neglect spin,   we shall just use eq. (\ref{eq:ab}) and we shall
not refer explicitly to any relativistic wave equation.

Indeed, things are different when a particle with spin moving in a
rotating frame is considered. In this case a coupling between the
spin and the angular velocity of the frame appears (this effect is
evaluated by Hehl-Ni\cite{hehl90}, Mashhoon\cite{mashhoon88} and
Papini\cite{papinilibro}).

Hence, the formal analogy that we are going to outline between
matter waves moving in a uniformly rotating frame, and charged
beams moving in a region where a constant magnetic potential is
present, holds only when the spin-rotation coupling is neglected.

\subsection{The relative space of a rotating disk}\label{ssec:relspace}

Before going on with our "demonstration by analogy", we want to
recall the definition of the "relative space" of a rotating disk,
that we introduced elsewhere\cite{rizzi02}. Since our analogy with
the Aharonov-Bohm effect is based on the measurements  performed
by the observers on the disk, the concept of relative space is
necessary to define, in a  proper  mathematical way, the physical
context in which  measurements are made. Even though a global
\textit{isotropic} 1+3 splitting of the space-time is not possible
when we deal with rotating observers (see Appendix A.\theCiOtto \
and A.\theCiQuattordici), the introduction of the relative space
allows well defined procedures for the space and time measurements
that can be performed (at large) by the observers in rotating
frames, and which reduce to the standard space and time
measurements locally (that is, in every Einstein-synchronized
LCIF). Let us outline the
main points that lead to the definition of the relative space.\\

\normalsize

The world-lines of each point of the rotating disk  are time-like
helixes (whose pitch, depending  on $\Omega $, is constant),
wrapping around the cylindrical surface  $r~=~const$, with $r \in
[0,R]$. These helixes fill, without intersecting, the whole
space-time region defined by $r\leq R<c/\Omega $;  they
constitute a time-like congruence $\Gamma$ which defines the rotating frame $K_{rot}$%
, at rest with respect to the disk.\footnote{%
\smallskip The constraint $R<c/\Omega $ simply means that the velocity of
the points of the disk cannot reach the speed of light. } Let us
introduce the coordinate transformation
\begin{equation}
\left\{
\begin{array}{rcl}
x'^{0} & = & ct^{\prime }=ct \\
x'^{1} & = & r^{\prime }=r \\
x'^{2} & = & \vartheta ^{\prime }=\vartheta -\Omega t \\
x'^{3} & = & z^{\prime }=z
\end{array}
\right. \mathrm{\ .}  \label{catt}
\end{equation}

The coordinate transformation $\left\{ x^{\mu }\right\}
\rightarrow \left\{ x'^{\mu }\right\}$ defined by (\ref{catt}) has
a kinematical meaning, namely it defines the passage from a chart
adapted to the inertial frame $K$ to a chart adapted to the
rotating frame $K_{rot}$. In the chart $\left\{ x'^{\mu}\right\}$
the metric tensor is written in the form:\footnote{For the sake of
simplicity, we substitute $r'=r$, from (\ref{catt})
$_{\mathit{II}}$.}
\begin{equation}
g_{\mu \nu }^{\prime }=\left(
\begin{array}{cccc}
-1+\frac{\Omega ^{2}{r}^{2}}{c^{2}} & 0 & \frac{\Omega {r}^{2}}{c} & 0 \\
0 & 1 & 0 & 0 \\
\frac{\Omega {r}^{2}}{c} & 0 & {r}^{2} & 0 \\
0 & 0 & 0 & 1
\end{array}
\right)  \label{born}
\end{equation}
This is the so called Born metric, and in the classic textbooks
(see, for instance Landau-Lifshits\cite{landau_ml} and M\o ller
\cite{moller}) it is commonly referred to as the space-time metric
in the rotating frame of the disk.

Moreover, we can calculate the space metric tensor $\gamma'_{ij}$
of the congruence which defines $K_{rot}$ (see Appendix
A.\theCiSei \ and A.\theCiQuattordici):
\begin{equation}
\gamma' _{ij}=g'_{ij}-\gamma' _{i}\gamma' _{j}=\left(
\begin{array}{ccc}
1 & 0 & 0 \\
0 & \frac{{r}^{2}}{1-\frac{\Omega ^{2}{r}^{2}}{c^{2}}} & 0 \\
0 & 0 & 1
\end{array}
\right) \mathrm{\ \quad .}  \label{eq:metricagamma}
\end{equation}

As it is shown explicitly  in  Appendix A.\theCiQuattordici, the
congruence $\Gamma$ of time-like helixes, wrapping around the
cylindrical hypersurfaces $\sigma _{r}\,$ ($r=cost\in ]0,R]$),
defines a Killing field not
in the Minkowski space-time $\mathcal{M}^{4}$, but in the submanifolds \textit{%
\ }$\sigma _{r}\;\subset \mathcal{M}^{4}$.

Consequently, we can point out the following interesting property.
Let $T_{p}=\Theta_p\oplus\Sigma_{p}$ be the tangent space to
$\mathcal{M}^4$ in $p$, where  $\Theta _{p}$, and $\Sigma _{p}$
are the  local time direction and the local space platform (see
Appendix A.\theCiSei). Then \textbf{\ }\textit{the splitting
}$T_{p}=\Theta _{p}\oplus \Sigma _{p}$\textit{\ and the space
metric tensor }$\gamma'_{ij}(p)$ are invariant along the lines of
$\Gamma$. Then it is  possible to define a one-parameter group of
diffeomorphisms with
respect to which both  the splitting $%
T_{p}=\Theta _{p}\oplus \Sigma _{p}$ and the space metric tensor
$\gamma'_{ij}(p)$ are invariant. The lines of $\Gamma$ constitute
the trajectories of this "space~$\oplus$~time" isometry. This
important property suggests a procedure to define an
\textit{extended} 3-space,  which we  call \textit{`relative
space'}: according to\cite{rizzi02}, it is recognized as the only
space having an actual physical meaning  from an operational point
of view, and it is identified as the 'physical space of a rotating
platform'. Let us briefly recall the definition of the relative
space. First of all, we introduce the following equivalence
relation between points and local space platforms:\\

RE\label{relequiv}:\textit{\ \noindent `` Two points (two local
space platforms) are equivalent if they belong to the same line of
the congruence $\Gamma$''}.\newline

The \textit{relative space} is the "quotient space" of the world
tube of the disk, with respect to the equivalence relation RE,
among points and local space platforms belonging to the lines of
the congruence $\Gamma$. In other words, each element of the
relative space is an equivalence class of \textit{points and local
space platforms} which verify the equivalence relation RE.\\

This definition simply means that the relative space is the
manifold whose ''points'' \textit{are} the lines of the
congruence. We point out that the presence of the local space
platforms in the equivalence relation RE is intended to emphasize
the vital role of the local Einstein synchronization associated to
each point of the relative space. Moreover we stress that it is
not possible to describe the relative space in terms of space-time
foliation, i.e. in the form $x^{0}=const$, where $x^{0}$ is an
appropriate coordinate time, because the space of the disk, as we
show in  Appendix A.\theCiQuattordici, is not time-orthogonal.
Hence, thinking of the space of the disk as a sub-manifold or a
subspace embedded in  space-time is misleading and meaningless.
The best we can do, if we long for some kind of visualization, is
to think of the relative space as the union of the infinitesimal
space platforms, each of which is associated, by means of the
request of $M$-orthogonality, to one and only one line of the
congruence.\\

In the relative space, an observer can perform measurements of
space and time. His reference frame, defined by the relative
space, coincides \textit{everywhere} with the local rest frame of
the rotating disk. As a consequence, space measurements are
performed on the bases of the spatial metric
(\ref{eq:metricagamma}), without caring of time, since
$\gamma'_{ij}$  does not depend on time.\footnote{This is a
consequence of the stationarity of the rotating frame. However, in
order to get rid of any misunderstandings (see for
instance\cite{klauber},\cite{klauberlibro},\cite{tartaglia_ml}),
we stress again that "without caring of time" does not mean
"without caring of synchronization". As a matter of fact, if
synchronization is not taken into account, rotation itself is not
taken into account.} Moreover, the observer can measure time
intervals using his own standard clock, on which he reads the
proper time.

\subsection{The Sagnac effect in the relative space}\label{ssec:ab2sagnac}

Now, we are going to describe the interference process of material
beams counter-propagating in a rotating ring interferometer, from
the viewpoint of the rotating frame. As we showed before, the
physical space of the rotating frame \textit{is} the relative
space. Then, a formal analogy between matter beams,
counter-propagating in the rotating frame, and charged beams,
propagating in a region where a magnetic potential is present, can
be outlined on the bases of Cattaneo's formulation of the
"standard relative dynamics". In particular, the equation of
motion of a particle relative to the rotating frame $K_{rot}$ can
be given in terms of the Gravitoelectromagnetic (GEM) fields (see
Appendix A.\theCiTredici). The introduction of the GEM fields
leads to an analogy between the Aharonov-Bohm effect and the
Sagnac effect in a fully relativistic context.

In eq. (\ref{eq:motogenF}), the general form of the standard
relative equation of motion of a particle is given in terms of the
gravito-electric field $\widetilde{\mathbf{E}}_G$, the
gravito-magnetic field $\widetilde{\mathbf{B}}_G$ and the external
fields.\footnote{For instance, the constraints that force the
particle to move along the rim of the disk are "external fields".}
In particular, in eq. (\ref{eq:motogenF}) a gravito-magnetic
Lorentz force appears
\begin{equation}
\widetilde{{\cal F}}_{i}=m\gamma _{0}\left( \frac{{\bf
\widetilde{v}}}{c}\times {\bf \widetilde{B}}_{G}\right) _{i}
\label{eq:genlorentz11}
\end{equation}

On the bases of this description,  we want to apply the formal
analogy between the gravito-magnetic and magnetic field to the
phase shift induced by rotation on a beam of massive particles
which, after being split, propagate in two opposite directions
along the rim of a rotating disk. When they are recombined, the
resulting phase shift is the manifestation of the Sagnac effect.

To this end, let us consider the analogue of the phase shift
(\ref{eq:ab}) for the gravito-magnetic field
\begin{equation}
\Delta \Phi =\frac{2m\gamma _{0}}{c\hbar }\oint_{C}{\bf \widetilde{A}}^{G}\cdot {\rm d}%
{\bf r}=\frac{2m\gamma _{0}}{c\hbar }\int_{S}{\bf
\widetilde{B}}^{G}\cdot {\rm d}{\bf S} \label{eq:gab}
\end{equation}
which is obtained on the bases of the formal analogy between eq.
(\ref{eq:genlorentz11}) and  the  magnetic force
(\ref{eq:lorentz1}):
\begin{equation}
\frac{e}{c}\mathbf{B} \rightarrow \frac{m \gamma_0}{c}
\mathbf{\widetilde{B}}^G \label{eq:sub2}
\end{equation}

To calculate the phase shift (\ref{eq:gab}) we must explicitly
express  the gravito-magnetic potential and field corresponding to
the congruence $\Gamma$  relative to the rotating frame $K_{rot}$.
In particular (see Appendix A.\theCiQuattordici) the non null
components of the vector field \mbox{\boldmath $\gamma$}$(x)$,
evaluated on the trajectory $R=const$ along which both beams
propagate, are:
\begin{equation}
\left\{
\begin{array}{c}
\gamma ^{0} \doteq \frac{1}{\sqrt{-g_{{00}}}} = \gamma  \\
\gamma _{0} \doteq \sqrt{-g_{{00}}} = \gamma ^{-1} \\
\gamma_2 = \gamma _{\vartheta}\doteq g_{\vartheta 0} \gamma
^{{0}}=\frac{\gamma \Omega R^{2}}{c}
\end{array}
\right. \label{eq:gammas}
\end{equation}
where $\gamma =\left( 1-\frac{\Omega ^{2}R^{2}}{c^{2}}\right) ^{-\frac{1}{2}%
} $.

As to the gravitomagnetic potential, we then obtain
\begin{equation}
\widetilde{A}_{2 }^{G}=\widetilde{A}_{\vartheta }^{G} \doteq
c^{2}\frac{\gamma _{\vartheta}}{\gamma _{0}} =\gamma ^{2}\Omega
R^{2}c \label{eq:vectpot}
\end{equation}

As a consequence, the phase shift (\ref{eq:gab}) becomes
\begin{equation}
\Delta \Phi =\frac{2m}{c\hbar \gamma }\int_{0}^{2\pi
}\widetilde{A}_{\vartheta }^{G}d\vartheta =\frac{2m}{c\hbar \gamma
}\int_{0}^{2\pi }\left( \gamma ^{2}\Omega R^{2}c\right) d\vartheta
=4\pi \frac{m}{\hbar }\Omega R^{2}\gamma \label{eq:deltaphigab}
\end{equation}

According to Cattaneo's terminology ( Appendix A.\theCiDieci), the
proper time is the "standard relative time" for an observer
on the rotating platform;  the proper time difference corresponding to (%
\ref{eq:deltaphigab}) is obtained according to

\begin{equation}
\Delta \tau = \frac{\Delta \Phi}{\omega}= \frac{\hbar}{E} \Delta
\Phi = \frac{\hbar}{mc^2}\Delta \Phi \label{eq:dtaudphi}
\end{equation}
and it turns out to be
\begin{equation}
\Delta \tau =4\pi \frac{\Omega R^{2}\gamma }{c^{2}} \equiv
\frac{4\pi R^2 \Omega}{c^2}\left( 1-\frac{\Omega ^{2}R^{2}}{c^{2}}
\right)^{-1/2} \label{eq:deltatau}
\end{equation}
Eq. (\ref{eq:deltatau}) agrees with the proper time difference
(\ref{eq:deltatausagnac}) due to the Sagnac effect, which, as we
pointed out in Subsection \ref{ssec:relativistic}, corresponds to
the time difference for any kind of matter entities
counter-propagating in a uniformly rotating disk. As we stressed
before,  this time difference does
not depend on the  standard relative velocity of the particles and it is exactly twice the {\it time lag }%
due to the synchronization gap arising in a rotating frame.\\

\textbf{Remark \ } In order to generalize Sakurai's procedure,
which refers to neutron beams, in this section we always referred
to material beams. However, the procedure that leads to the time
difference (\ref{eq:deltatau}) can be carried out also for light
beams. Actually, in Appendix A.\theCiDodici, we show that a
standard relative
 mass of a photon $m\doteq\frac{h\nu}{c^2}$ can be consistently defined.
Consequently, the relative formulation of the equation of motion
of a photon is described in a way analogous to that of a material
particle, and the procedure that we have just outlined can be
applied in a straightforward way to massless particles too.\\

The phase shift (\ref{eq:deltaphigab}) can be expressed also as a
function of the area $S$ of the surface enclosed by the
trajectories:
\begin{equation}
\Delta \Phi = 2 \beta^2 S \Omega \frac{m}{\hbar} \frac{\gamma^2}{\gamma-1}= 2%
\frac{m}{\hbar} S \Omega \left (\gamma+1 \right)
\label{eq:phiarea}
\end{equation}
where $\beta = \frac{\Omega R}{c}$ and
\begin{equation}
S=\int_0^R \int_0^{2\pi} \frac{r dr d\vartheta}{\sqrt{1-\frac{\Omega^2 r^2}{%
c^2}}}=2\pi \frac{c^2}{\Omega^2}\left( 1-\sqrt{1-\frac{\Omega^2 R^2}{c^2}}%
\right) =2\pi \frac{c^2}{\Omega^2}\left(\frac{\gamma-1}{\gamma}
\right) \label{eq:area1}
\end{equation}
We notice that (\ref{eq:phiarea}) reduces to
(\ref{eq:phaseomega1})\footnote{Apart a factor 2, whose origin has
been explained in the footnote \fnref{fn:fact2} in Subsection
\ref{ssec:tests}.} only in the first order approximation with
respect to $\frac{\Omega R}{c}$ (when $\gamma \rightarrow 1$): the
formal difference between (\ref{eq:phiarea}) and
(\ref{eq:phaseomega1}) is due to the non Euclidean features of the
relative space (Appendix A.\theCiQuattordici).

\section{Conclusions}\label{sec:conclusions}

The relativistic Sagnac effect has been deduced by means of two
derivations.\\

In the first part of this paper a direct derivation has been
outlined on the bases of the relativistic kinematics. In
particular, only the law of velocity addition has been used to
obtain the Sagnac time difference,  and to show, in a
straightforward way, its independence from the physical nature and
the velocities (relative to the
turntable) of the interfering beams.\\

In the second part of this paper, an alternative derivation has
been presented. In particular, the formal analogy outlined by
Sakurai, which explains the effect of rotation using a
"ill-assorted" mixture of non-relativistic quantum mechanics and
Newtonian mechanics (which are Galilei-covariant), and
intrinsically relativistic elements\footnote{Indeed, the lack of
self-consistency, due to the use of this "odd mixture", is present
not only in Sakurai's derivation, but also in all  known
approaches based on the formal analogy with the Aharonov-Bohm
effect.} (which are Lorentz-covariant), has been extended to a
fully relativistic treatment, using the 1+3 Cattaneo's splitting
technique. The space in which the waves propagate has been
recognized as the relative space of a rotating frame, which turns
out to be non-Euclidean.  In this way, we have obtained   a
derivation of the relativistic Sagnac time difference (whose first
order approximation coincides with Sakurai's result)
in a self-consistent way.\\

Both derivations are carried out in a fully relativistic context,
which turns out to be the natural arena where the Sagnac effect
can be explained. Indeed, its universality can be clearly
understood as a purely geometrical effect in the Minkowski
space-time of the SRT, while it is hard to grasp in the context of
classical physics.

\appendix
%\chapappendix{Space-Time Splitting and Cattaneo's
%Approach}\label{sec:cattaneo}

\section{Appendix: Space-Time Splitting and Cattaneo's Approach}\label{sec:cattaneo}

The tools for splitting space-time have had a great (even though
heterogenous) development in the years, and they have been used in
various application in General Relativity(GR). Indeed, the common
aim of the different approaches to splitting techniques is the
description of what is measured by a test family of observers,
moving along certain curves in the four-dimensional continuum.

In this way, locally, along the world-lines of these observers,
space+time measurements can be recovered, and the description of
the physical phenomena borrowed from the SRT can be transferred
into GR.

There are various approach to splitting of space-time, and a great
work has be done, recently, to describe everything in a common
framework, by stressing the connections among the different
techniques\cite{jantzen01a},\cite{jantzen92},\cite{bini00}.

Probably, the most well known and used splitting is the so called
"ADM splitting"\cite{ADM} (see also
\textit{Gravitation}\cite{MTW}), which is based on the use of a
family of space-like hypersurfaces ("slicing" point of view); on
the other hand, the approach based on a congruence of time-like
observers ("threading" point of view) was developed independently
by various authors, such as Cattaneo\cite{cattaneo}, \cite{catt1},
\cite{catt2}, \cite{catt3}, \cite{catt4}, M\o ller\cite{moller}
and Zel'manonv\cite{{zelmanov56}} during the 1950's, but it has
remained greatly unknown for a long time, also because some of the
original works were not published in English, but in Italian,
French or Russian. Because of the pedagogical aim that we have in
writing this paper, we decided to present here a very introductory
primer to the original Cattaneo's works on splitting of
space-time. After the publication of his works during the 1950's
and 1960's, a lot of work has been done, in order to improve his
techniques (see \cite{jantzen01a},\cite{jantzen92},\cite{bini00},
and references therein). However, we believe that the foundations
of Cattaneo's approach can be understood, in the easiest and most
enlightening way, by referring to his original works. Moreover,
since our aim is not historical but pedagogical, we shall
translate his "relative formulation of dynamics" in terms of the
Gravitoelectromagnetic analogy: indeed, we exploited this analogy
in our derivation of the Sagnac effect starting from the
Aharonov-Bohm effect.\\

%\indent \setcounter{subsection}{1} \textbf{\thesubsection \
%}\label{ssec:c1}

\indent \textbf{A.\theCiUno \ }\label{ssec:c1} It is important to
define correctly the properties of the physical frames with
respect to which we describe the measurement processes. We shall
adopt the most general description, which takes into account
non-inertial frames (for instance rotating frames) in the SRT, and
arbitrary frames in GR.

The physical space-time is a (pseudo)riemannian manifold $%
\mathcal{M}^{4}$, that is a pair $\left(
\mathcal{M},\mathbf{g}\right)$, where $\mathcal{M}$ is a connected 4-dimensional Haussdorf manifold and $%
\mathbf{g}$ is the metric tensor.\footnote{%
The riemannian structure implies that $\mathcal{M}$ is endowed
with an affine connection compatible with the metric, i.e. the
standard Levi-Civita connection.} Let the signature of the
manifold be $(-1,1,1,1)$. Suitable differentiability conditions,
on both $\mathcal{M}$ and $\mathbf{g}$, are assumed.

\indent \textbf{A.\theCiDue \ }\label{ssec:c2_ml}A physical
reference frame  is a time-like congruence
 $\Gamma $: the set of the world lines of the test-particles
constituting the ''reference fluid''.\footnote{The concept of
'congruence' refers to a set of word lines filling the manifold,
or some part of it, smoothly, continuously and without
intersecting. The concept of 'reference fluid' is an obvious
generalization of the 'reference solid' which can be used in flat
space-time, where the test particles constitute a global inertial
frame. In this case, their relative distance remains constant and
they evolve as a rigid frame. However: (i) in GR  test particles
can be subject to a gravitational field (curvature of space-time);
(ii) in the SRT test particles can be subject to an acceleration
field. In both cases, global inertiality is lost and tidal effects
arise, causing a variation of the distance between them. So we
must speak of "reference fluid", dropping the compelling request
of classical rigidity.} The congruence $\Gamma $ is identified by
the field of unit vectors tangent to its world lines. Briefly
speaking, the congruence is the (history of the) physical frame or
the reference fluid (they are synonymous).\newline

\indent \textbf{A.\theCiTre \ }\label{ssec:c3}Let $\{x^{\mu
}\}=(x^{{0}},x^{1},x^{2},x^{3})$ be a system of coordinates in the
neighborhood of a point $p\in \mathcal{M}$; these coordinates are
said to be \textit{admissible} (with respect to the
congruence $\Gamma $) when\footnote{%
Greek indices run from 0 to 3, Latin indices run from 1 to 3.}
\begin{equation}
g_{{00}}<0\ \ \ \ g_{ij }dx^{i }dx^{j }>0 \label{eq:admiss}
\end{equation}
Thus the coordinates $x^{{0}}=var$ can be seen as describing the
world lines of the $\infty ^{3}$ particles of the reference
fluid.\\

\indent \textbf{A.\theCiQuattro \ }\label{ssec:c4}When a reference
frame has been chosen, together with a set of admissible
coordinates, the most general coordinates transformation which
does not change the physical frame, i.e. the congruence $\Gamma $,
has the form (see \cite{moller},\cite{gron1},\cite{cattaneo}):
\begin{equation}
\left\{
\begin{array}{c}
x^{\prime }{}^{{0}}=x^{\prime }{}^{{0}}(x^{{0}},x^{1},x^{2},x^{3}) \\
x^{\prime }{}^{i}=x^{\prime }{}^{i}(x^{1},x^{2},x^{3})
\end{array}
\right.  \label{eq:gauge_trans}
\end{equation}
with the additional condition $\partial x^{\prime 0}/\partial
x^{{0}}>0$, which  ensures that the  change of time
parameterization does not change the arrow of time. The
coordinates transformation (\ref
{eq:gauge_trans}) is said to be \textit{internal to the physical frame} $%
\Gamma $, or   \textit{internal gauge transformation}, or more
simply \textit{Cattaneo's  gauge
transformation}.\\

\indent \textbf{A.\theCiCinque \ }\label{ssec:c5}An ``observable''
physical quantity is in general frame-dependent, but its physical
meaning requires that it cannot depend on the particular
parameterization of the physical frame: in brief, it cannot be
gauge-dependent. Then a problem arises. In the mathematical model
of GR, physical quantities are expressed by absolute
entities,\footnote{ `Absolute' means `independent of any reference
frame'.} such as world-tensors, and the physical laws, according
to the covariance principle, are just relations among these
entities. So, given a reference frame, how do we relate these
absolute quantities to the relative, i.e. reference-dependent,
ones? And how do we relate world equations to reference-dependent
ones? In other words: how do we relate, by a suitable 1+3
splitting, the mathematical model of space-time to the
observable quantities which are relative to a reference frame?\\

\indent \textbf{A.\theCiSei \ }\label{ssec:c6} In order to do
that, we are going to introduce the \textit{projection technique}
developed by Cattaneo. Let \mbox{\boldmath $\gamma$}$(x)$ be the
field of unit vectors tangent to the world lines of the congruence
$\Gamma $. Given a time-like congruence $\Gamma $ it is always
possible to choose a system of admissible
coordinate so that the lines $x^{0}=var$ coincide with the lines of $\Gamma $%
; in this case, such coordinates are said to be \textit{`adapted
to the physical frame'} defined by the congruence $\Gamma $.

Being $g_{\mu \nu }\gamma ^{\mu }\gamma ^{\nu }=-1$, we get
\begin{equation}
\left\{
\begin{array}{c}
\gamma ^{{0}}=\frac{1}{\sqrt{-g_{{00}}}} \\
\gamma ^{i}=0
\end{array}
\right. \;\;\;\;\;\;\;\;\;\;\;\;\left\{
\begin{array}{c}
\gamma _{0}=\sqrt{-g_{{00}}} \\
\gamma _{i}=g_{i{0}}\gamma ^{{0}}
\end{array}
\right.  \label{eq:gammas11}
\end{equation}

In each point $p\in \mathcal{M}$, the tangent space $T_{p}$ can be
split into the direct sum of two subspaces: $\Theta _{p}$, spanned
by $\gamma ^{\alpha }$, which we shall call \textit{local time
direction} of the given frame, and $\Sigma _{p}$, the
3-dimensional subspace which is supplementary (orthogonal) with
respect to $\Theta_{p}$; $\Sigma _{p}$ is called \textit{local
space platform} of the given frame. So the tangent space can be
written as the direct sum
\begin{equation}
T_{p}=\Theta _{p}\oplus \Sigma _{p}  \label{eq:tangsum}
\end{equation}

A vector $\mathbf{v}\in T_{p}$ can be projected onto $\Theta _{p}$ and $%
\Sigma _{p}$ using the \textit{time projector} $\gamma _{\mu
}\gamma _{\nu }$ and the\textit{\ space projector }$\gamma _{\mu
\nu }\doteq g_{\mu \nu }-\gamma _{\mu }\gamma _{\nu }$:
\begin{equation}
\left\{
\begin{array}{rclll}
\bar{v}_{\mu } & = & P_{\Theta }\left( \,\mathbf{v}\right) &
\doteq &
\gamma _{\mu }\gamma _{\nu }v^{\nu } \\
\widetilde{v}_{\mu } & = & P_{\Sigma }\left( \,\mathbf{v}\right) &
\doteq & v_{\mu }-v^{\nu }\gamma _{\nu }\gamma _{\mu }=\left(
g_{\mu \nu }-\gamma _{\mu }\gamma _{\nu }\right) v^{\nu }=\gamma
_{\mu \nu }v^{\nu }
\end{array}
\right.  \label{eq:DefSplitComp}
\end{equation}

\textbf{Notation} The superscripts $^{-},^{\sim }$ denote
respectively a \textit{time vector} and a \textit{space vector},
or more generally, a \textit{time tensor }and\textit{\ a space
tensor} (see below).\\

Equation (\ref{eq:DefSplitComp}) defines the \textit{natural
splitting} of a vector. The tensors $\gamma _{\mu }\gamma _{\nu }$
and $\gamma _{\mu \nu }$ are called \textit{time metric tensor}
and \textit{space metric tensor}, respectively. In particular, for
each vector $\mathbf{v}$ it is possible to define a `time norm'
$\left\| \,\mathbf{v\,}\right\| _{\Theta }$ and a `space norm'
$\left\| \,\mathbf{v\,}\right\| _{\Sigma }$ as follows:
\begin{equation}
\left\| \,\mathbf{v}\,\right\| _{\Theta } \doteq \bar{v}_{\rho
}\bar{v}^{\rho }=\gamma _{\rho }\gamma _{\nu }v^{\nu }\left(
\gamma ^{\rho }\gamma _{\mu }v^{\mu }\right) =\gamma _{\mu }\gamma
_{\nu }v^{\mu }v^{\nu }=(\gamma _{\mu }v^{\mu })^{2} \leq 0
\label{eq:normtemp}
\end{equation}
\begin{equation}
\left\| \,\mathbf{v}\,\right\| _{\Sigma } \doteq \widetilde{v}_{\nu }\widetilde{v}%
^{\nu }= \gamma _{\mu \nu }v^{\mu }(v^{\nu }-\gamma _{\eta }\gamma
^{\nu }v^{\eta })=\gamma _{\mu \nu }v^{\mu }v^{\nu } \geq 0
\label{eq:normspaz}
\end{equation}

For a tensor field $T\in T_{p}$, every index can be projected onto
$\Theta _{p}$ and $\Sigma _{p}$ by means of the projectors defined
before:
\begin{equation}
P_{\Theta }(T_{...\mu ...}) \doteq \gamma _{\mu }\gamma _{\nu
}T^{...\nu ...}\ \ \ P_{\Sigma }(T_{...\mu ...}) \doteq \gamma
_{\mu \nu }T^{...\nu ...}  \label{eq:protn}
\end{equation}

A tensor field of order two can be split into the sum of four
tensors
\begin{equation}
\begin{array}{rclcrcl}
P_{\Sigma \Sigma }\,\left( \,{t}_{\mu \nu }\,\right) & \doteq &
\gamma _{\mu \rho
}\gamma _{\nu \eta }t^{\rho \eta } & \;\quad & P_{\Sigma \Theta }\,\left( \,{%
t}_{\mu \nu }\,\right) & \doteq & \gamma _{\mu \rho }\gamma _{\nu
}\gamma _{\eta
}t^{\rho \eta } \\
P_{\Theta \Sigma }\,\left( \,{t}_{\mu \nu }\,\right) & \doteq &
\gamma _{\mu }\gamma _{\rho }\gamma _{\nu \eta }t^{\rho \eta } &
\quad & P_{\Theta \Theta }\,\left( \,{t}_{\mu \nu }\,\right) &
\doteq & \gamma _{\mu }\gamma _{\nu }\gamma _{\rho }\gamma _{\eta
}t^{\rho \eta }
\end{array}
\label{eq:ProiettoriR2}
\end{equation}
belonging to four orthogonal subspaces
\begin{equation}
T_{p}\otimes T_{p}=(\Sigma _{p}\otimes \Sigma _{p}) \oplus (\Sigma
_{p}\otimes \Theta _{p}) \oplus (\Theta _{p}\otimes \Sigma _{p})
\oplus (\Theta _{p}\otimes \Theta _{p} )  \label{eq:Sottospazi}
\end{equation}

In particular, every tensor belonging entirely to $\Sigma
_{p}\otimes \Sigma
_{p}$ is called a \textit{space tensor} and every tensor belonging to $%
\Theta _{p}\otimes \Theta _{p}$ is called a \textit{time tensor}.
Of course, these entities have a tensorial behavior only with
respect to the group of the coordinates transformation
(\ref{eq:gauge_trans}). It is straightforward to extend these
procedures and definitions to tensors of generic order $n$ (see
below) .\newline

\textbf{Remark 1 }The natural splitting of a tensor is
gauge-independent: it depends only on the physical frame chosen.
The projection technique gives  gauge-invariant quantities that
have an operative meaning in our physical frame; namely, they
represent the objects of our measurements.
\newline

\textbf{Remark 2 } In $\Gamma$-adapted coordinates, a \textit{time
vector} $\mathbf{\bar{v}\in } \ \Theta _{p}$ is
characterized by the vanishing of its controvariant space components ($\bar{v%
}^{i}=0$); a \textit{space vector} $\mathbf{\tilde{v}\in } \
\Sigma _{p}$ by the vanishing of its covariant time component
($\tilde{v}_{0}=0$). As a generalization: (i) a given index of a
tensor $\mathbf{T}$ is called a \textit{time-index} if all the
tensor components of the type $T_{...}^{.i.}$ ($i \in [1,2,3]$)
vanish; (ii) a given index of a tensor $\mathbf{T}$ is called a
\textit{space-index} if all the tensor components of the type
$T_{.0.}^{...}$ vanish. For a \textit{time tensor}, i.e. for a
tensor belonging to $\Theta_p \ \otimes...\otimes\Theta_p$,
property (i) holds for all its indices; for a \textit{space
tensor}, i.e. a tensor belonging to $\Sigma_p \ \otimes ...\otimes
\Sigma_p$, property (ii)
holds for all its indices.\\

\indent \textbf{A.\theCiSette \ }\label{ssec:c7}To formulate the physical equations relative to the frame $%
\Gamma $, we need the following differential operator
\begin{equation}
\tilde{\partial}_{\mu }\doteq \partial _{\mu }-\gamma _{\mu }\gamma ^{{0}%
}\partial _{{0}}  \label{eq:dertras}
\end{equation}
which is called \textit{transverse partial derivative}. It is a
"space vector" and (its definition) is gauge-invariant.\newline

It is easy to show that, for a generic scalar field $\varphi (x)$
we obtain:
\begin{equation}
P_{\Sigma }(\partial _{\mu }\varphi )=\tilde{\partial}_{\mu
}\varphi \label{eq:protrans}
\end{equation}
So $\tilde{\partial}_{\mu }$ defines the \textit{transverse
gradient,} i.e. the space projection of the local
gradient.\newline

The projection technique that we have just outlined allows to
calculate the projections of the Christoffel symbols. It is
remarkable that the total space projections turn out to be
\begin{eqnarray}
P_{\Sigma \Sigma \Sigma }(\mu \nu ,\lambda ) &=&\frac{1}{2}\left( \tilde{%
\partial _{\mu }}\gamma _{\nu \lambda }+\tilde{\partial _{\nu }}\gamma
_{\lambda \mu }-\tilde{\partial _{\lambda }}\gamma _{\mu \nu
}\right) \doteq
\widetilde{\Gamma}^{*}_{\mu \nu \lambda}  \nonumber \\
P_{\Sigma \Sigma \Sigma }\left\{ _{\mu \nu }^{\lambda }\right\} &=&%
\widetilde{\Gamma}^{*}_{\mu \nu \sigma}\gamma ^{\sigma \lambda }
\doteq \widetilde{\Gamma}^{* \ \lambda}_{\mu \nu}
\label{eq:prochris}
\end{eqnarray}
where the space metric tensor $\gamma _{\mu \nu}$ substitutes the metric tensor $%
g_{\mu \nu }$ and the transverse derivative substitutes the
''ordinary'' partial derivative.\newline

\indent \textbf{A.\theCiOtto \ }\label{ssec:c8}  The differential
features of the congruence $\Gamma $ are described by the
following tensors
\begin{eqnarray}
\widetilde{C}_\mu &=&\gamma ^{\nu }\nabla _{\nu }\gamma _{\mu }
\label{eq:cmu} \\
\widetilde{\Omega }_{\mu \nu } &=&\gamma _{0}\left[ \widetilde{\partial }%
_{\mu }\left( \frac{\gamma _{\nu }}{\gamma _{0}}\right)
-\widetilde{\partial }_{\nu }\left( \frac{\gamma _{\mu }}{\gamma
_{0}}\right) \right]
\label{eq:vortex} \\
\widetilde{K}_{\mu \nu } &=&\gamma ^{{0}}\partial _{{0}}\gamma
_{\mu \nu } \label{eq:born}
\end{eqnarray}
$\widetilde{C }_\mu $ is the \textit{curvature vector}, $\widetilde{\Omega }%
_{\mu \nu }$ is the \textit{space vortex tensor}, which gives the
local angular velocity of the reference fluid, $\widetilde{K}_{\mu
\nu }$ is the \textit{Born space tensor}, which gives the
deformation rate of the reference fluid; when this tensor is null,
the frame is said to be rigid according to the definition of
rigidity given by Born\cite{born}.

In a relativistic context the classical concept of rigidity, which
is dynamical in its origin, since it is based on the presence of
forces that are responsible for rigidity, becomes meaningless.
Born's definition of rigidity is the natural generalization of the
classical one. It depends on the motion of the test particles of
the congruence: hence, it is a kinematical constraint. According
to Born, a body moves rigidly if the space distance
$\sqrt{\gamma_{ij}dx^i dx^j}$ between neighbouring points of the
body, as measured in their successive (locally inertial) rest
frames, is constant in time.\footnote{In the simple case of
translatory motion, a body moves rigidly if, at every moment, it
has a Lorentz contraction corresponding to its observed
instantaneous velocity, as measured by an inertial observer.} For
Born's condition of rigidity see also
Rosen\cite{rosen}, Boyer\cite{boyer}, Pauli\cite{pauli}.\\

\textbf{Definitions} The following definitions\footnote{%
It is worthwhile to notice that, in the literature, there is not
common agreement about these definitions, see for instance
Landau-Lifshits\cite{landau_ml}.} are referred to the (geometry of
the) physical frame $\Gamma$:

\begin{itemize}
\item  \textit{constant} - when there exists at least one adapted
chart, in which the components of the metric tensor are not
depending on the time coordinate: $\partial _{0}g_{\mu \nu }=0$

\item  \textit{time-orthogonal} - when there exist at least one
adapted chart in which $g_{0i}=0$; in this system the lines
$x^{{0}}=var$ are orthogonal to the 3-manifold $x^{{0}}=cost$

\item  \textit{static} -  when there exists at least one adapted chart in which $%
g_{0i}=0$ and $\partial _{0}g_{\mu \nu }=0$.

\item  \textit{stationary} when it is constant and non
time-orthogonal
\end{itemize}

\textbf{Remark 3 }The condition of being time-orthogonal is a
property of the physical frames, and not of the coordinate
systems: for a reference frame to be time-orthogonal it is
necessary and sufficient that the space vortex
$\widetilde{\Omega}_{\mu\nu}$
tensor vanishes.\\

When the space vortex tensor is null, moreover, the fluid is said
to be irrotational;  if both the curvature vector and the space
vortex tensor are zero, the fluid is said irrotational and
geodesic. When the space vortex tensor is not null, a global
synchronization of the standard clocks in the frame is not
possible.\\

The irrotational, rigid and geodesic motion (of a frame) is
characterized by the condition $\nabla _{\mu }\gamma _{\nu }=0$:
this is the generalization, in a curved space-time context, of the
translational uniform motion in flat space-time.\newline

\indent \textbf{A.\theCiNove \ }\label{ssec:c9} The natural
splitting also permits  to calculate the   Riemann curvature
tensor of the 3-space of the reference frame. The complete space
projection of the curvature tensor of space-time is
\cite{cattaneo}:
\begin{equation}
P_{\Sigma \Sigma \Sigma \Sigma}\left(R_{\mu \nu \sigma \rho} \right)=%
\widetilde{R}_{\mu \nu \sigma \rho}^{*}-\frac{1}{4}\left(\tilde{\Omega}%
_{\sigma \mu} \tilde{\Omega}_{\rho \nu}-\tilde{\Omega}_{\rho \mu}\tilde{%
\Omega}_{\sigma \nu} \right)-\frac{1}{2} \tilde{\Omega}_{\mu \nu} \tilde{%
\Omega}_{\sigma \rho}   \label{eq:rie_space11}
\end{equation}
where
\begin{eqnarray}
\widetilde{R}_{\mu \nu \rho \sigma }^{*} &\doteq
&\frac{1}{4}\left(
\widetilde{\partial }_{\rho \mu }\gamma _{\nu \sigma }-\widetilde{\partial }%
_{\sigma \mu }\gamma _{\nu \rho }+\widetilde{\partial }_{\sigma
\nu }\gamma
_{\mu \rho }-\widetilde{\partial }_{\rho \nu }\gamma _{\mu \sigma }\right) +%
\frac{1}{4}\left( \widetilde{\partial }_{\mu \rho }\gamma _{\nu
\sigma
}-\right.  \nonumber \\
&&\left. \widetilde{\partial }_{\nu \rho }\gamma _{\mu \sigma }+\widetilde{%
\partial }_{\nu \sigma }\gamma _{\mu \rho }-\widetilde{\partial }_{\mu
\sigma }\gamma _{\nu \rho }\right) +\gamma ^{\alpha \beta }\left[ \widetilde{%
\Gamma }_{\sigma \nu ,\alpha }^{*}\widetilde{\Gamma }_{\rho \mu ,\beta }^{*}-%
\widetilde{\Gamma }_{\rho \nu ,\alpha }^{*}\widetilde{\Gamma
}_{\sigma \mu ,\beta }^{*}\right] \nonumber \\ & &
\label{eq:curvspaz}
\end{eqnarray}
The space Christoffel symbols are defined in eq.
(\ref{eq:prochris}). Since it has all space indices (see Remark 2,
Subsection A.\theCiSei ), the curvature tensor (\ref{eq:curvspaz})
is a space tensor. Then the curvature tensor which is adequate to
describe the space geometry of the physical frame $\Gamma$ is the
space part $\widetilde{R}_{ijkl}^{*}$ of the tensor
(\ref{eq:curvspaz}).

In particular, if we deal with flat space-time, since the
curvature tensor $R_{\mu\nu \sigma \rho}$ is null, from
(\ref{eq:rie_space11}) we get
\begin{equation}
P_{\Sigma \Sigma \Sigma \Sigma}\left(R_{\mu \nu \sigma \rho} \right)=%
\widetilde{R}_{\mu \nu \sigma \rho}^{*}-\frac{1}{4}\left(\tilde{\Omega}%
_{\sigma \mu} \tilde{\Omega}_{\rho \nu}-\tilde{\Omega}_{\rho \mu}\tilde{%
\Omega}_{\sigma \nu} \right)-\frac{1}{2} \tilde{\Omega}_{\mu \nu} \tilde{%
\Omega}_{\sigma \rho}=0   \label{eq:rie_space22}
\end{equation}
Eq. (\ref{eq:rie_space22}) shows that, in this case, the space
components $\widetilde{ R}_{ijkl }^{*}$ are completely defined by
the terms containing the space vortex tensor, which is related to
rotation: hence, the non Euclidean nature of the space of a
rotating frame depends only on  rotation itself.\\

\indent \textbf{A.\theCiDieci \ }\label{ssec:c10}  Let us consider
two infinitesimally close events in space-time, whose coordinates
are $x^\alpha$ and $x^\alpha+dx^\alpha$. We can
introduce the following definitions:\\
\textit{"standard relative time"}\\
\begin{equation}
dT=-\frac{1}{c}\gamma_\alpha dx^\alpha \label{eq:dTdef}
\end{equation}
\textit{"standard relative space element"}
\begin{equation}
d\sigma^2=\gamma_{\alpha\beta}dx^\alpha dx^\beta \equiv
\gamma_{ij}dx^idx^j \label{eq:dsigmadef}
\end{equation}
It is evident that these quantities are strictly dependent on the
physical frame defined by the vector field \mbox{\boldmath
$\gamma$}. They have a fundamental role in the standard relative
formulation of the kinematics and dynamics of a particle in an
inertial or gravitational field. To this end, it is worthwhile to
notice that both $dT$ and $d\sigma$ are invariant with respect to
the internal gauge transformations (\ref{eq:gauge_trans}). More
generally speaking, all  laws of relative kinematics and dynamics
that we are going to illustrate, are invariant with respect to
(\ref{eq:gauge_trans}): in other words, their formulation will
depend only on the choice of the congruence $\Gamma$, and it will
be independent of the (adapted) coordinates chosen to parameterize
the physical frame defined by $\Gamma$.

By using (\ref{eq:dTdef}) and (\ref{eq:dsigmadef}) it is easy to
show that the space-time invariant $ds^2$ can be written in the
form
\begin{equation}
ds^2=d\sigma^2-c^2dT^2 \label{eq:dsdsgimadT}
\end{equation}
Let us consider the motion of a point in $\mathcal{M}^4$. The
world-line of a material particle is time-like ($ds^2<0$), while
it is light-like ($ds^2=0$) for a photon (or for a generic
massless particle) . The following definition applies to a
particle $P$ in the physical frame $\Gamma$: $P$ is at rest if its
world line coincides with one of the lines of the congruence. In
other words, $dP\parallel {\boldmath \gamma}$ and $dx^i \equiv 0$.

On the contrary, when the world-line of the point $P$ does not
coincide with any of the lines of $\Gamma$, the point is said to
be in motion in the given physical frame. Since  $dx^i \neq 0$, we
can write the parametric equation of the world-line of $P$ in
terms of a parameter $\lambda$, $x^\alpha=x^\alpha(\lambda)$; $dP$
is either time-like or light-like and in both cases $dT\neq 0$, so
we can express the world-coordinates of the moving particle using
the standard relative time as a parameter:
$x^\alpha=x^\alpha(T)$.\\

\textbf{Remark \ } From the very definition (\ref{eq:dTdef}), it
is evident that $dT$ represents the proper time measured by an
observer at rest ($dx^i=0$) in $\Gamma$.\\

\indent \textbf{A.\theCiUndici \ }\label{ssec:c11} Let
$v^\alpha=\frac{dx^\alpha}{dT}$ be the relative 4-velocity. We
shall call \textit{"standard relative velocity"} its spatial
projection
\begin{equation}
\widetilde{v}_\beta \doteq P_\Sigma
(v_\beta)=\gamma_{\beta\alpha}\frac{dx^\alpha}{dT}=\gamma_{\beta
i}\frac{dx^i}{dT} \label{eq:relvel1}
\end{equation}
Since $\widetilde{v}_\beta \in \Sigma_p$, then
$\widetilde{v}_0=0$. The controvariant components of the standard
relative velocity are
\begin{equation}
\widetilde{v}^i=\frac{dx^i}{dT} \ \ \ \widetilde{v}^0=-\gamma_i
\frac{\widetilde{v}^i}{\gamma_0} \label{eq:relvel2}
\end{equation}
(because $\gamma_\alpha \widetilde{v}^\alpha=0$).

As a consequence, eq. (\ref{eq:relvel1}) can be written as
\begin{equation}
\widetilde{v}_i=\gamma_{ij}\widetilde{v}^j \label{eq:relvel3}
\end{equation}
The (space) norm of the standard relative velocity is (see eq.
(\ref{eq:normspaz}))
\begin{equation}
\left\| \,\mathbf{v}\,\right\| _{\Sigma }\doteq
\widetilde{v}^2=\gamma_{ij}\widetilde{v}^i{}\widetilde{v}^j=\frac{d\sigma^2}{dT^2}
\label{eq:relvel4}
\end{equation}
In particular, for a photon, since $ds^2=0$, we get from eq.
(\ref{eq:dsdsgimadT}( $\widetilde{v}^2=c^2$, which is the same
result that one would expect in the SRT. Dealing with material
particles, we can introduce the proper time
$d\tau^2=-\frac{1}{c^2}d\sigma^2$, and, using (\ref{eq:dTdef}), we
can write
\begin{equation}
\frac{d\sigma^2}{dT^2}=-c^2 \frac{d\tau^2}{dT^2}
\label{eq:gammal1}
\end{equation}
Taking into account (\ref{eq:relvel4}) we obtain
\begin{equation}
\frac{dT}{d\tau}=\frac{1}{\sqrt{1-\frac{\widetilde{v}^2}{c^2}}}
\label{eq:gammal2}
\end{equation}
This relation is formally identical to the  one that is valid in
the SRT. By using the definitions of  standard relative time
(\ref{eq:dTdef}) and standard relative velocity
(\ref{eq:relvel1}), it is possible to obtain the following
relation between $dT$ and the coordinate time interval
$dt=\frac{dx^0}{c}$:
\begin{equation}
dT\left(1+\gamma_i \frac{\widetilde{v}^i}{c} \right)=-\gamma_0dt
\label{eq:dTdt}
\end{equation}

Summarizing, we have shown that Cattaneo's projection technique,
endowed with the standard relative quantities defined before,
allows to  extend formally the laws of the SRT to any physical
reference frame, in presence of gravitational or inertial fields.\\

\textbf{Remark \ } In general, the standard relative time that we
have introduced is not an exact differential: this means that, in
a generic frame $\Gamma$ we cannot define a unique standard time,
or, in other words, the global synchronization of the standards
clocks is not possible. In order to have a globally well defined
standard relative time, the $\gamma_\alpha$ must be identified as
the partial derivatives of a scalar function $f$:
$\gamma_\alpha=\partial_\alpha f$, and this is possible iff
$\Omega_{\alpha\beta} \equiv \partial_\alpha
\gamma_\beta-\partial_\beta \gamma_\alpha=0$, that is when the
physical frame is both irrotational (i.e.
$\widetilde{\Omega}_{\alpha\beta}=0$) and geodesic(i.e.
$\widetilde{C}_\alpha=0$).\footnote{It is easy to verify that
$\Omega_{\alpha\beta}=\widetilde{\Omega}_{\alpha\beta}+\widetilde{C}_\alpha
\gamma_\beta-\gamma_\alpha \widetilde{C}_\beta$.}\\

\indent \textbf{A.\theCiDodici \ }\label{ssec:c12} The equation of
motion of a free  mass point is a geodesic of the differential
manifold $\mathcal{M}^4$, endowed with the Levi-Civita connection.
The connection coefficients, in the coordinates $\{x^\mu\}$
adapted to the physical frame are $\Gamma^{\alpha}_{\
\beta\gamma}$. Explicitly, the geodesic equations is written as
\begin{equation}
\frac{DU^\alpha}{d\tau}=0 \Leftrightarrow
\frac{dU^\alpha}{d\tau}+\Gamma^\alpha_{\ \beta\gamma}U^\beta
U^\gamma=0 \label{eq:geo1}
\end{equation}
in terms of the 4-velocity $U^\alpha$ and the proper time $\tau$.
Let $m_0$ be the proper mass of the particle: then the
energy-momentum 4-vector is $P^{\alpha}=m_0U^{\alpha}$. We can
write the geodesic equation also in the covariant and
contravariant forms
\begin{equation}
\frac{DP_{\alpha}}{d\tau}=0 \ \ \ \frac{DP^{\alpha}}{d\tau}=0
\label{eq:geo2}
\end{equation}
or, equivalently, using the standard relative time
\begin{equation}
\frac{DP_{\alpha}}{dT}=0 \ \ \ \frac{DP^{\alpha}}{dT}=0
\label{eq:geoT}
\end{equation}
Now we want to re-formulate the geodesic equation in its relative
form, i.e. by means of the standard relative quantities that we
have introduced so far. To this end, let us define the
\textit{standard relative momentum}
\begin{equation}
\widetilde{p}_\alpha \doteq P_\Sigma \left( P_\alpha
\right)=\gamma_{\alpha\beta}P^\beta=m_0\gamma_{\alpha
 i }\frac{dx^i}{dT}\frac{dT}{d\tau}=m\widetilde{v}_\alpha
\label{eq:relmom1}
\end{equation}
where the \textit{standard relative mass}
\begin{equation}
m\doteq\frac{m_0}{\sqrt{1-\frac{\widetilde{v}^2}{c^2}}}
\label{eq:relmass}
\end{equation}
has been introduced, in formal analogy with the SRT. Since
$\widetilde{p}_\alpha \in \Sigma_p$, then $\widetilde{p}_0=0$.

We can also define the \textit{standard relative energy}
\begin{equation}
E\doteq-c\gamma_\alpha P^\alpha=-m_0 c \gamma_i
\frac{dx^i}{d\tau}=m_0c^2\frac{dT}{d\tau}=mc^2 \label{eq:relen1}
\end{equation}
recovering the well known relation which is used in the SRT.
Notice also that
\begin{equation}
P_\Theta \left( P_\alpha \right) = \frac{E}{c}\gamma_\alpha
\label{eq:relen2}
\end{equation}

For a massless particle, like a photon, we can define the
energy-momentum 4-vector
\begin{equation}
P^\alpha=\frac{h\nu}{c^2}\frac{dx^\alpha}{dT}
\label{eq:palphafotone}
\end{equation}
where $h$ is the Planck constant and, in terms of relative
quantities the relation that links the  wavelength and the
frequency of the photon to the velocity of light is
$\lambda\nu=\frac{d\sigma}{dT}=c$.

So, for a photon , we can introduce the \textit{standard relative
energy}
\begin{equation}
E=-c\gamma^\alpha P_\alpha \doteq h\nu \label{eq:enfotone}
\end{equation}
the \textit{standard relative  mass}
\begin{equation}
m \equiv \frac{E}{c^2}=\frac{h\nu}{c^2} \label{eq:mfotone}
\end{equation}
and the \textit{standard relative momentum}
\begin{equation}
\widetilde{p}_\alpha \equiv m \widetilde{v}_\alpha
\label{eq:pfotone}
\end{equation}

The equation of motion of a free photon is a null geodesic
\begin{equation}
\frac{D P_\alpha}{dT}=0 \label{eq:geofotone}
\end{equation}
where the standard relative time has been used to parameterize it.

The spatial projection of the  geodesic equations for matter
(\ref{eq:geoT}) and light-like particles (\ref{eq:geofotone}) is
written in the form
\begin{equation}
P_\Sigma \left( \frac{DP_\alpha}{dT}\right) \equiv
\frac{\hat{D}\widetilde{p}_\alpha}{dT}-m\widetilde{G}_\alpha=0
\Leftrightarrow
\frac{\hat{D}\widetilde{p}_i}{dT}-m\widetilde{G}_i=0
\label{eq:geospace1}
\end{equation}
where
\begin{equation}
\frac{\hat{D}\widetilde{p}_i}{dT} \doteq
\frac{d\widetilde{p}_i}{dT}-\widetilde{\left(ij, k
\right)^*}\widetilde{p}^k \frac{dx^j}{dT} \label{eq:dhat}
\end{equation}
and
\begin{equation}
\widetilde{G}_i=-c^2\widetilde{C}_i+c\widetilde{\Omega}_{ij}\widetilde{v}^j
\label{eq:gi}
\end{equation}

Hence, we can write the space projection of the geodesic equation
in the simple form
\begin{equation}
\frac{\hat{D}\widetilde{p}_i}{dT}=m\widetilde{G}_i
\label{eq:geospace2}
\end{equation}
where it is shown that the variation of the spatial momentum
vector is determined by the field $\widetilde{G}_i$.

It is often useful to split the field $\widetilde{G}_i$ into the
sum of two fields $\widetilde{G'_i},\widetilde{G''_i}$, defined as
follows:
\begin{eqnarray}
\widetilde{G'}_i    & \doteq &
-c^2\widetilde{C}_i=-c^2\left(\gamma_0 \tilde{\partial}_i
\gamma^0-\partial_0\left(\frac{\gamma_i}{\gamma_0}\right) \right)  \nonumber \\
\widetilde{G''}_i   & \doteq & c \widetilde{\Omega}_{ij}
\widetilde{v}^j \label{eq:gprimogsecondo}
\end{eqnarray}
The field $\widetilde{G'}_i$ can be interpreted as a dragging
gravitational-inertial field ($c^2C_\alpha$ is the 4-acceleration
$a_\alpha$ of the particle of the reference frame) and the field
$\widetilde{G''}_i$ can be interpreted as a Coriolis-like
gravitational-inertial field. Actually, starting from the space
vortex tensor of the congruence
\begin{equation}
\widetilde{\Omega}_{hk}\doteq \gamma _{0}\left[ \tilde{\partial}_{h}\left( \frac{%
\gamma _{k}}{\gamma _{0}}\right) -\tilde{\partial}_{k}\left(
\frac{\gamma _{h}}{\gamma _{0}}\right) \right]
\label{eq:vorticespaziale}
\end{equation}
we can introduce  $\widetilde{\mbox{\boldmath $\omega$}}$ $(x)$ $
\in \Sigma _{p}$, which is the axial 3-vector associated to
$\widetilde{\Omega}_{hk}$, by means of the relation
\begin{equation}
\widetilde{\omega} ^{i}\doteq \frac{c}{4}\varepsilon ^{ijk}\tilde{\Omega}_{jk}=\frac{c}{2%
}\varepsilon ^{ijk}\gamma _{o}\tilde{\partial}_{j}\left( \frac{\gamma _{k}}{%
\gamma _{o}}\right)   \label{eq:vettorevortice}
\end{equation}
where $\epsilon^{ijk} \doteq \frac{1}{\sqrt{det(\gamma
_{ij})}}\delta ^{ijk}$ is the Ricci-Levi Civita tensor, defined in
terms of the completely antisymmetric permutation symbol $\delta
^{ijk}$ and of the spatial metric tensor $\gamma_{ij}$. As a
consequence, we can write $\widetilde{G}''_i$ in the form
\begin{equation}
\widetilde{G}''_{i}=2m(\widetilde{{\bf v}}\times
\widetilde{\mbox{\boldmath $\omega$}})_{i} \label{eq:corgen}
\end{equation}
which corresponds to a generalized Coriolis-like force.  So, the
equation of motion (\ref{eq:geospace2}) can be written in the form
\begin{equation}
\frac{\hat{D}\widetilde{\mathbf{p}}}{dT}=-m\widetilde{\mathbf{a}}+2m(\widetilde{{\bf
v}}\times \widetilde{\mbox{\boldmath $\omega$}})
\label{eq:inertial}
\end{equation}
where $\widetilde{\mathbf{a}}$ is the spatial projection of the
4-acceleration $a_\alpha$.

From (\ref{eq:inertial}) we see that the relative formulation of
the equation of motion of a free particle is identical to the
expression of the classical equation of motion of a particle which
is acted upon by inertial fields only. Moreover, if $m$ is defined
by eq. (\ref{eq:mfotone}) the equation of motion
(\ref{eq:inertial}) holds also for massless
particles.\\

\indent \textbf{A.\theCiTredici \ }\label{ssec:c13} Now let us
turn back to eq. (\ref{eq:geospace2}). We can introduce the
''gravito-electric potential'' $\phi ^{G}$ and the
''gravito-magnetic potential'' $\widetilde{A}_{i}^{G}$ defined by
\begin{equation}
\left\{
\begin{array}{c}
\phi ^{G}\doteq -c^{2}\gamma ^{0} \\
\widetilde{A}_{i}^{G}\doteq c^{2}\frac{\gamma _{i}}{\gamma _{0}}
\end{array}
\right.   \label{eq:gengravpot11}
\end{equation}

As we shall see in a while, these names are justified by the fact
that, introducing the "Gravitoelectromagnetic" (GEM) potentials
and fields, eq. (\ref{eq:geospace2}) can be written as the
equation of motion of a particle under the action of a generalized
Lorentz
force.\\

In terms of these potentials, the vortex 3-vector
$\widetilde{\omega}^i$ is expressed in the form

\begin{equation}
\widetilde{\omega} ^{i}=\frac{1}{2c}\varepsilon ^{ijk}\gamma _{0}\left( \tilde{\partial}%
_{j}\widetilde{A}_{k}^{G}\right)   \label{eq:omega1}
\end{equation}
and, by introducing  the ''gravito-magnetic field"
\begin{equation}
\widetilde{B}_{G}^{i}\doteq \left( \widetilde{{\bf \nabla }}\times
{\bf \widetilde{A}}_{G}\right) ^{i}  \label{eq:gengravmag}
\end{equation}
eq. (\ref{eq:omega1})   can be written as
\begin{equation}
\widetilde{\omega}^{i}= \frac{1}{2c}\gamma
_{0}\widetilde{B}_{G}^{i} \label{eq:omega2}
\end{equation}

As a consequence, the velocity-dependent force (\ref{eq:corgen})
becomes
\begin{equation}
\widetilde{G}''_{i}=m\gamma _{0}\left( \frac{{\bf
\widetilde{v}}}{c}\times {\bf \widetilde{B}}_{G}\right) _{i}
\label{eq:genlorentz}
\end{equation}

Moreover, the dragging term $\widetilde{G}_{i}^{\prime }$ (see eq.
\ref{eq:gprimogsecondo})
\begin{equation}
\widetilde{G}_{i}^{\prime }=-\left( -\widetilde{\partial} _{i}\phi
_{G}-\partial _{0}\widetilde{A}^{G}_i\right)
\end{equation}
can be interpreted as a "gravito-electric field":
\begin{equation}
\widetilde{E}^{G}_i\doteq -\left( -\widetilde{\partial} _{i}\phi
_{G}-\partial _{0}\widetilde{A}^{G}_i\right)  \label{eq:egen}
\end{equation}

Then, the equation of motion (\ref{eq:geospace2}) can be written
in the form
\begin{equation}
\frac{\hat{D}\widetilde{p}_{i}}{dT}=m   \widetilde{E}^{G}_i+m\gamma_0\left( \frac{{\bf \widetilde{v}}}{c}\times {\bf \widetilde{B}}%
_{G}\right) _{i}  \label{eq:motogen1}
\end{equation}
which looks like the equation of motion of a particle acted upon
by a "generalized" Lorentz force.\\

If the particle is not free, its equation of motion is
\begin{equation}
\frac{DP^{\alpha }}{dT }=F^{\alpha } \label{eq:eqmoto4d}
\end{equation}
where the external field  is described by the 4-vector $F^{\alpha
}$. The space projection of (\ref{eq:eqmoto4d}) then becomes
\begin{equation}
\frac{\hat{D}\widetilde{p}_{i}}{dT}=m   \widetilde{E}^{G}_i+m\gamma_0\left( \frac{{\bf \widetilde{v}}}{c}\times {\bf \widetilde{B}}%
_{G}\right) _{i} + \widetilde{F}_i \label{eq:motogenF}
\end{equation}
where the space projection $\widetilde{F}_i$ of the external field
has been introduced.\\

\textbf{Remark 1 } As we have just outlined, the splitting in
curved space-time leads to a non-linear analogy with
electromagnetism in flat space-time, which is commonly referred to
as "Gravitoelectromagnetism"\cite{jantzen01a}. Namely, the local
fields, due to the "inertial forces" felt by the test observers,
are associated to Maxwell-like fields: in particular, a
gravito-electric field is associated to the local linear
acceleration, while a gravito-magnetic field is associated to
local angular acceleration (that is, to local
rotation).\footnote{This analogy, built in fully non linear GR, in
its linear approximation corresponds to the well known analogy
between the theory of electromagnetism and the linearized theory
of General
Relativity\cite{mashhoon01},\cite{ruggiero02}.}\\

\textbf{Remark 2 } We want to point out that while the field
$\widetilde{G}_i$ is  gauge invariant, its components
$\widetilde{G}''_i$ and $\widetilde{G}''_i$ are not separately
gauge invariant. In other words, the gravito-electric field
$\widetilde{E}^{G}_i$ and gravito-magnetic field
$\widetilde{B}^{G}_i$ are not invariant with respect to gauge
transformations (\ref{eq:gauge_trans}).\footnote{It can be shown
that they are invariant with respect to a smaller group of gauge
transformations. For instance, they are invariant with respect to
\begin{equation}
x'^0=ax^0+b \ \ \ \ \ \ \ \ \ \ \ \ \ x'^i=x'^i(x^1,x^2,x^3)
\label{eq:spatem}
\end{equation}
where $a,b$ are constants.}\\

 \indent \textbf{A.\theCiQuattordici \ }\label{ssec:c14} In this subsection, a great number of
calculations are explicitly carried out. The geometric objects
which we deal with, always refer to the physical frame $K_{rot}$;
the lines of the congruence which identifies this physical frame,
are described in Subsection \ref{ssec:relspace}, and the passage
from the inertial frame $K$ to the rotating frame $K_{rot}$ is
defined by the coordinates transformations $\left\{ x^{\mu
}\right\} \rightarrow \left\{ x'^{\mu }\right\}$
\begin{equation}
\left\{
\begin{array}{rcl}
x'^{0} & = & ct^{\prime }=ct \\
x'^{1} & = & r^{\prime }=r \\
x'^{2} & = & \vartheta ^{\prime }=\vartheta -\Omega t \\
x'^{3} & = & z^{\prime }=z
\end{array}
\right. \mathrm{\ .}  \label{catt1}
\end{equation}

However, here and henceforth, for the sake of simplicity, we shall
not use primed letters. In particular, all  space vectors belong
to the (tangent bundle to the)  relative
space of the disk, which has been defined in Subsection \ref{ssec:relspace}.\\

The metric tensor expressed in coordinates adapted to the rotating
frame is
\begin{equation}
g_{\mu \nu }=\left(
\begin{array}{cccc}
-1+\frac{\Omega ^{2}{r}^{2}}{c^{2}} & 0 & \frac{\Omega {r}^{2}}{c} & 0 \\
0 & 1 & 0 & 0 \\
\frac{\Omega {r}^{2}}{c} & 0 & {r}^{2} & 0 \\
0 & 0 & 0 & 1
\end{array}
\right)  \label{born1}
\end{equation}
and its contravariant components  are:
\begin{equation}
g^{\mu \nu }=\left(
\begin{array}{cccc}
-1 & 0 & \frac{\Omega }{c} & 0 \\
0 & 1 & 0 & 0 \\
\frac{\Omega }{c} & 0 & \frac{1-\frac{\Omega
^{2}{r}^{2}}{c^{2}}}{r^{2}} &
0 \\
0 & 0 & 0 & 1
\end{array}
\right)  \label{bornup}
\end{equation}

The non zero Christoffel symbols turn out to be
\begin{eqnarray}
\Gamma _{00}^{\ \ 1} &=&-\frac{\Omega ^{2}}{c^{2}}r  \label{christoffel} \\
\Gamma _{01}^{\ \ 2 } &=&\frac{\Omega }{cr}  \nonumber \\
\Gamma _{02 }^{\ \ 1} &=&-\frac{\Omega }{c}r  \nonumber \\
\Gamma _{12 }^{\ \ 2 } &=&\frac{1}{r}  \nonumber \\
\Gamma _{22 }^{\ \ 1} &=&-r  \nonumber
\end{eqnarray}
The non null components of the vector field \mbox{\boldmath
$\gamma$}$(x)$ are:
\begin{equation}
\left\{
\begin{array}{c}
\gamma ^{0} \doteq \frac{1}{\sqrt{-g_{{00}}}} =
\frac{1}{\sqrt{1-\frac{\Omega^2 r^2}{c^2}}}
  \\
\gamma _{0} \doteq \sqrt{-g_{{00}}} = {\sqrt{1-\frac{\Omega^2 r^2}{c^2}}} \\
\gamma _{2}\doteq g_{2 0} \gamma ^{{0}}=\frac{ \Omega r^{2}}{c}
\frac{1}{\sqrt{1-\frac{\Omega^2 r^2}{c^2}}}
\end{array}
\right. \label{eq:gammas111}
\end{equation}

and the components of the space metric tensor are turn out to be
\begin{equation}
\gamma _{ij}=g_{ij}-\gamma _{i}\gamma _{j}=\left(
\begin{array}{ccc}
1 & 0 & 0 \\
0 & \frac{{r}^{2}}{1-\frac{\Omega ^{2}{r}^{2}}{c^{2}}} & 0 \\
0 & 0 & 1
\end{array}
\right) \mathrm{\ \quad .}  \label{eq:metricagamma11}
\end{equation}

The  non null components of the space vortex tensor are:
\begin{equation}
\widetilde{\Omega}_{12}=\gamma_0 \tilde{\partial_1}\left(\frac{\gamma_2}{%
\gamma_0} \right) = \left( \sqrt{1-\frac{\Omega ^{2}r^{2}}{c^{2}}}%
\right) ^{-3}\frac{%
2\Omega r}{c} \label{eq:omega12all}
\end{equation}
As a consequence, the rotating frame is not time orthogonal.

Moreover, the spatial Born tensor is null:
\begin{equation}
\widetilde{K}_{ij}\doteq \gamma^0 \partial_{0}\gamma_{ij}=0
\label{eq:kappaij}
\end{equation}
since the space metric (\ref{eq:metricagamma11}) does not depend
on the time coordinate. Hence the rotating frame $K_{rot}$ is
rigid, in the sense of  Born's definition of
rigidity (Section A.\theCiOtto ).\\

The covariant components of the Killing tensor  of the congruence
$\Gamma$ turn out to be
\begin{equation}
K_{\mu \nu}\equiv \gamma _{\mu ;\nu }+\gamma _{\nu ;\mu
}=\frac{\partial \gamma
_{\mu }}{\partial x^{\nu }}+\frac{\partial \gamma _{\nu }}{\partial x^{\mu }}%
-2\Gamma _{\mu \nu }^{\ \ \alpha }\gamma _{\alpha }
\label{eq:killinall}
\end{equation}
Taking into account (\ref{christoffel}) and (\ref{eq:gammas111}),
we obtain that the only non null components in $\mathcal{M}^{4}$
are:
\begin{equation}
K_{01}\equiv \gamma _{0;1}+\gamma _{1;0}=\frac{\partial \gamma _{0}}{%
\partial r}-2\Gamma _{01}^{\ \ 2 }\gamma _{2 }=\frac{\partial
\gamma _{0}}{\partial r}-g^{2 \alpha }\frac{\partial g_{0\alpha }}{%
\partial r}  \label{eq:terre}
\end{equation}
\begin{equation}
K_{2 1}\equiv \gamma _{2 ;1}+\gamma _{1;2 }=\frac{\partial \gamma
_{2 }}{\partial r}-2\Gamma _{2 1}^{\ \ 2 }\gamma
_{2 }=\frac{\partial \gamma _{2 }}{\partial r}-g^{2 \alpha }%
\frac{\partial g_{2 \alpha }}{\partial r} \label{eq:thetaerre}
\end{equation}

The components $K_{01}$, $K_{21}$ depend solely on the partial
derivatives with respect to $r$ of some functions of $r$. If we
evaluate these components in $\mathcal{M}^{4}$, we obtain a non
zero result, while if we evaluate the same components on the
cylindrical hypersurface $\sigma _{r} \equiv \{ r=const\;(>0)\}$,
they result identically zero. Summing up, we get:
\begin{equation}
K_{01}\equiv \gamma _{0;1}+\gamma _{1;0}=\left\{
\begin{array}{c}
-\frac{\Omega ^{2}r}{c^{2}}\left( \sqrt{1-\frac{\Omega ^{2}r^{2}}{c^{2}}}%
\right) ^{-1}\;\neq 0\;\mathit{\ in \mathit{\
}}\mathcal{M}^{4}\mathit{\ }
\\
0\;\mathit{in \ the \ submanifold \ }\sigma _{r}\;(r=const)
\end{array}
\right.  \label{eq:Ktr}
\end{equation}
\begin{equation}
K_{2 1}\equiv \gamma _{2 ;1}+\gamma _{1;2 }=\left\{
\begin{array}{c}
-\frac{\Omega ^{3}r^{3}}{c^{3}}\left( \sqrt{1-\frac{\Omega ^{2}r^{2}}{c^{2}}}%
\right) ^{-3}\neq 0\ \;\;\mathit{in \mathit{\ }}\mathcal{M}^{4} \\
0\;\mathit{in \ the \ submanifold \ }\sigma _{r}\;(r=const)
\end{array}
\right.  \label{eq:Kthetar}
\end{equation}

Equations  (\ref{eq:Ktr}) and  (\ref{eq:Kthetar}) show that the
time-like helixes congruence $\Gamma$ defines a Killing field in
the submanifold $\sigma _{r}\;\subset \mathcal{M}^{4}$ but this is
not
a  Killing field in $\mathcal{M}^{4}$.\\

We get the same result if we express the Killing tensor  using the
Born tensor $\widetilde{K}_{\mu \nu }$ and the curvature vectors
$\widetilde{C}_{\nu }$ of the lines of the congruence $\Gamma \,$:
\[
K_{\mu \nu }\equiv \gamma _{\mu ;\nu }+\gamma _{\nu ;\mu }=\widetilde{K}%
_{\mu \nu }+\gamma _{\mu }\widetilde{C}_{\nu }+\widetilde{C}_{\mu
}\gamma _{\nu }
\]
For the rotating disk ($\widetilde{K}_{\mu \nu }=0$) we simply
obtain:
\begin{equation}
K_{\mu \nu }=\gamma _{\mu }\widetilde{C}_{\nu }+\widetilde{C}_{\mu
}\gamma _{\nu }  \label{bb}
\end{equation}

Equation (\ref{bb}) is very interesting, because it shows the
geometrical meaning of the fact that the Killing tensor is zero in
the submanifold $\sigma _{r}\;\subset \mathcal{M}^{4}$, but it is
not zero in $\mathcal{M}^{4}$. In fact, the congruence $\Gamma$ of
time-like helixes is geodesic on $\sigma _{r}\,$ (where
$\widetilde{C}_{\mu }=0 $), but of course not in
$\mathcal{M}^{4}$,\footnote{Apart from the degenerate case $r=0$,
which corresponds to a straight line in $\mathcal{M}^{4}$.} where
the curvature vector $\tilde{C}_{\mu }=\gamma ^{\alpha }\nabla
_{\alpha }\gamma _{\mu }$ has the following non-null component:
\begin{equation}
\widetilde{C}_{1}=\gamma ^{0}\frac{\frac{\Omega ^{2}r}{c^{2}}}{\sqrt{1-\frac{%
\Omega ^{2}r^{2}}{c^{2}}}}=-\frac{\Omega ^{2}r}{c^{2}-\Omega
^{2}r^{2}} \label{eq:vettoredicurvatura}
\end{equation}
As a consequence, the Killing tensor has the following non-null
components:

\begin{equation}
K_{01}=\gamma _{0}\widetilde{C}_{1}=-\frac{\Omega ^{2}r}{c^{2}}\left( \sqrt{1-%
\frac{\Omega ^{2}r^{2}}{c^{2}}}\right) ^{-1}  \label{eq:Ktr1}
\end{equation}
\begin{equation}
K_{2 1}=\gamma _{2 }\widetilde{C}_{1}=-\frac{\Omega ^{3}r^{3}}{%
c^{3}}\left( \sqrt{1-\frac{\Omega ^{2}r^{2}}{c^{2}}}\right) ^{-3}
\label{eq:Kthetar1}
\end{equation}
Equations (\ref{eq:Ktr1}) and  (\ref{eq:Kthetar1}) are in
agreement, respectively, with equations (\ref{eq:Ktr}) and
(\ref{eq:Kthetar}).

The only non zero components of the curvature tensor of space of
the disk are
\begin{equation}
\widetilde{R}_{1212}^{*}=-3\frac{\left( \frac{\Omega r}{c}\right) ^{2}}{%
\left( 1-\left( \frac{\Omega r}{c}\right) ^{2}\right)
^{3}}\label{riemann}
\end{equation}

and those which are obtained by the symmetries of the indices.\\
The non null components of the space projection of the Ricci
tensor, are:
\begin{equation}
\begin{array}{rcl}
\widetilde{R}_{11}^{*} & = & -3\frac{\Omega ^{2}}{c^{2}}\,\left(1-%
\frac{\Omega ^{2}r^{2}}{c^{2}} \right)^{-2} \\
\widetilde{R}_{22}^{*} & = & -3\,\frac{\Omega ^{2}}{c^{2}}\,\left(1-%
\frac{\Omega ^{2}r^{2}}{c^{2}} \right)^{-3}
\end{array}
\label{ricci}
\end{equation}
Finally the curvature scalar is:
\begin{equation}
\widetilde{R}^{*}=-6\,\frac{\Omega ^{2}}{c^{2}}\,\left(1-%
\frac{\Omega ^{2}r^{2}}{c^{2}} \right)^{-2} \label{scalar}
\end{equation}

The calculation of the curvature of the space of the rotating
disk,  which turns out to be non Euclidean (in particular
hyperbolic), confirms Einstein's early intuition\cite{einstein50}
about the relations between curvature and rotation.\footnote{See,
for instance, Stachel\cite{stachel}.}

\end{document}